\documentclass[draftclsnofoot, onecolumn]{IEEEtran}
\usepackage{amsmath,amsfonts}
\usepackage{algorithm}
\usepackage{algorithmic}
\usepackage{array}
\usepackage[caption=false,font=normalsize,labelfont=sf,textfont=sf]{subfig}
\usepackage{textcomp}
\usepackage{stfloats}
\usepackage{url}
\usepackage{cite}
\usepackage{subcaption}
\usepackage{verbatim}
\usepackage{tikz}
\usepackage{amssymb}
\usepackage{todonotes}
\usetikzlibrary{shapes, arrows,positioning,fit,backgrounds, calc}
\usepackage{chemformula}
\usepackage{graphicx}
\usepackage{lipsum}
\def\BibTeX{{\rm B\kern-.05em{\sc i\kern-.025em b}\kern-.08em
    T\kern-.1667em\lower.7ex\hbox{E}\kern-.125emX}}
\usepackage{balance}
\usepackage{authblk}
\begin{document}
\title{ Cold-Diffusion Driven Downward Continuation of Gravity Data}

\author{Adarsh Jain, Pawan Bharadwaj, and Chandra Sekhar Seelamantula \IEEEmembership{Senior Member, IEEE}
\thanks{
%Manuscript submitted on June 25, 2025; 

Adarsh Jain is with the IISc Mathematics Initiative (IMI), Department of Mathematics, Indian Institute of Science, Bengaluru, Karnataka 560012, India (e-mail: adarshjain1@iisc.ac.in). 

Pawan Bharadwaj is with the Centre for Earth Sciences, Indian Institute of Science, Bengaluru, Karnataka 560012, India (e-mail: pawan@iisc.ac.in).

Chandra Sekhar Seelamantula is with the Department of Electrical Engineering, Indian Institute of Science, Bengaluru, Karnataka 560012, India (e-mail: css@iisc.ac.in).}}

% \markboth{IEEE Transactions on Computational Imaging}%
% {}
\markboth{Preprint}%
{}

\maketitle

\begin{abstract}
Gravity data can be better interpreted after enhancing high-frequency information via downward continuation. Downward continuation is an ill-posed deconvolution problem.
It has been tackled using regularization techniques, which 
are sensitive to the choice of regularization parameters.
More recently, convolutional neural networks such as the U-Net have been trained using synthetic data to potentially learn prior information and perform deconvolution without the need to adjust the regularization parameters. Our experiments reveal that the U-Net is highly sensitive to  correlated noise, which is ubiquitously present in geophysical field data. In this paper, we develop a framework based on the \textbf{\emph{cold-diffusion model}} using the exponential kernel associated with downward continuation. The exponential form of the kernel allows us to train the U-Net to tackle multiple concurrent deconvolution problems with varying levels of blur. This allows our framework to be more robust and quantitatively outperform traditional U-Net-based approaches. The performances also closely matches that of \textbf{\emph{oracle}} Tikhonov reconstruction technique, which has access to the ground truth.
\end{abstract}


\section{Introduction}
\begin{figure}[t]
	\centering
	\includegraphics[width=4.25in]{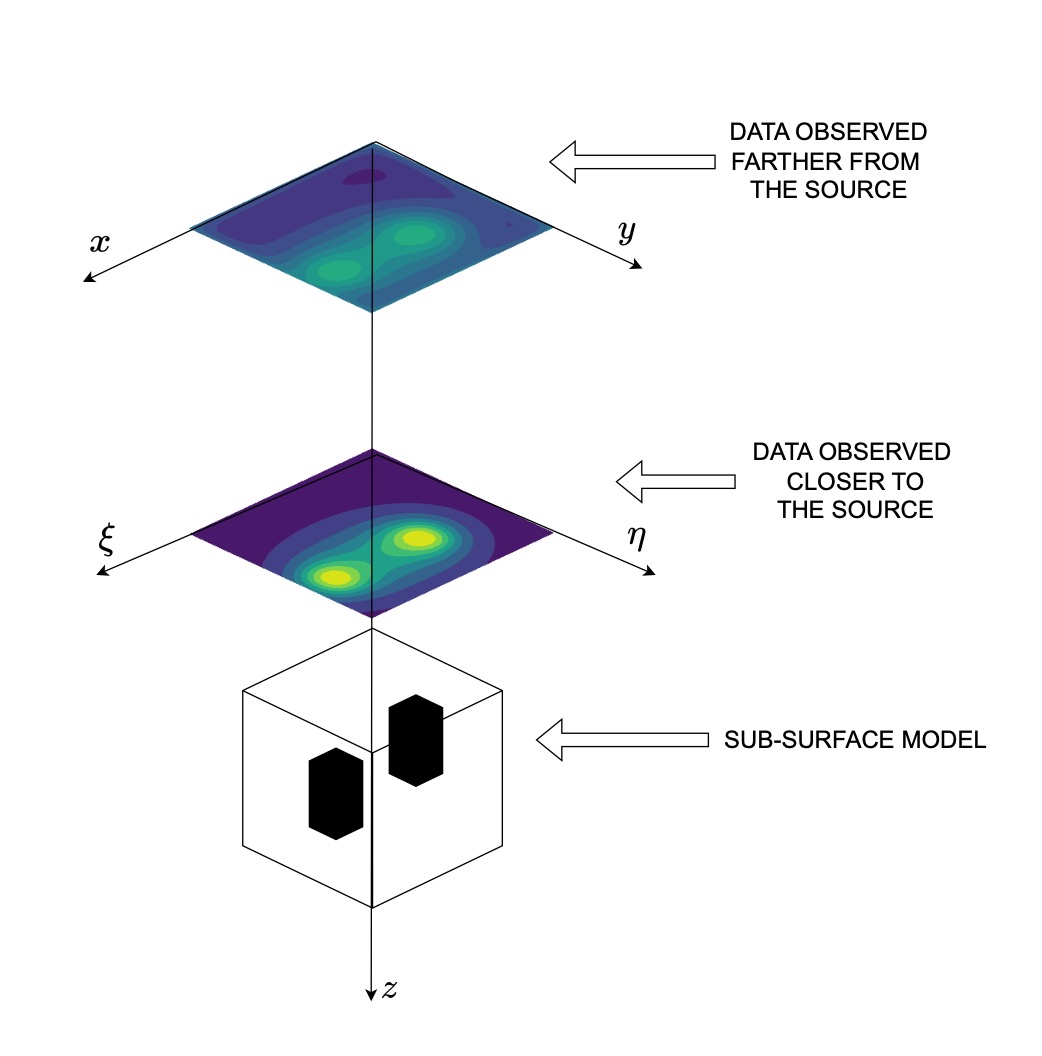}
	\caption{Schematic illustration of downward continuation.}
	\label{fig1}
\end{figure}

\IEEEPARstart{P}{otential} field data, such as gravity and magnetic data, are significant for mineral exploration tasks, as they provide critical information about the density and susceptibility of the subsurface, respectively. 
Numerous studies have leveraged gravity data for diverse applications, including mineral exploration, \ch{CO_2} plume detection, and subsurface imaging \cite{huangetal, Celaya2023, Liy2022, Xian2025}. Specifically, airborne gravity data has proven to be valuable for subsurface investigations, as demonstrated by Meng et al. \cite{Meng2021} and Wang et al. \cite{Wang2022}. Recent studies by Wilkinson et al. \cite{Wilkinson2017} and Kabirjadeh et al. \cite{kabirzadeh2017}  have shown the application of gravity and micro-gravity surveys in monitoring \ch{CO2} storage sites, which is an essential aspect of sustainability efforts.\\
\indent Downward continuation~\cite{Ze2021, wang23} is a technique that enhances the high-frequency signals related to subsurface anomalies, facilitating a more accurate interpretation of the potential field data for mineral exploration. This technique involves reconstructing the gravity data closer to the subsurface sources, based on observations or measurements taken from a plane located at a distance due to experimental limitations. With advances in geophysical data acquisition, the need for effective downward continuation (DC) has increased, especially with the increasing use of airborne surveys. Technologies such as unmanned aerial vehicles (UAVs), including survey aircraft and drones, offer cost-effective and efficient alternatives to traditional ground surveys; however, potential fields become smoother due to the increased distance from sources, making downward continuation a key technique for restoring the lost high frequencies in the airborne data~\cite{Zhao2018}.\\
\indent DC is an ill-posed deconvolution problem. Wang et al. \cite{wang23} reviewed  analytical methods for stable DC and concluded that Tikhonov regularized DC (TRDC) \cite{tikhonov} and truncated Taylor-series based iterative DC (TTSIDC) \cite{zhang} are the best performing techniques.
Downward continuation of magnetic data has been studied extensively, with Li et al. \cite{Li2012} applying it to enhance magnetic data for UXO application and Wang et al. \cite{Ze2021} introducing an iterative method that utilizes adaptive filtering. Motivated by these applications, this study focuses on devising a stable solution for the downward continuation of gravity data. Fig.~\ref{fig1} shows a schematic diagram of downward continuation.\\
\indent Deep learning has emerged as a prominent technique in recent years, particularly for applications such as image segmentation \cite{Minaee2021}, image generation \cite{elasri2022}, and solving ill-posed problems \cite{Adler2017}. Recently, numerous researchers \cite{Celaya2023,Liy2022, Xian2025, zhang2021} have utilized U-Net, self-supervised learning and diffusion models in gravity method for geophysical exploration. Diffusion models  \cite{cao2024} represent a significant branch within the various deep learning frameworks, specifically in generative modeling, originally developed for unconditional generation. Recently, Bansal et al. \cite{colddif} introduced the cold-diffusion network, showcasing the conditional generation capability of diffusion models. They demonstrated that these models can invert arbitrary transformations, effectively eliminating the need for Gaussian noise in the diffusion framework. We conceptualize the downward continuation problem as a deblurring problem, where the diffusion steps correspond to the degree of smoothness introduced to the data at different continuation heights. Our aim is to leverage the cold-diffusion-model for solving the inverse problem.

\subsection{Problem Formulation}
\indent Continuation refers to the process of transforming the data observed at one level to correspond to another level. Specifically, upward continuation involves deriving data corresponding to an elevation higher than the observation plane. For example, consider the ($x, y$) and ($\xi, \eta$) planes as shown in Fig.~\ref{fig1}, which are separated by a distance $\Delta z$. The ($\xi, \eta$) plane is positioned closer to the source. The mathematical expression for upward continuation, as formulated by Blackey \cite{blackey}, is presented as follows:
\begin{equation}
	\label{eqn-1}
	U(x, y, \Delta z) = \frac{\Delta z}{2 \pi}\iint\frac{V{(\xi, \eta, 0)}\,\mathrm{d}\xi \,\mathrm{d}\eta}{\left((x - \xi)^2+(y - \eta)^2 + {\Delta z}^2\right)^{\frac{3}{2}}},
\end{equation}
 where the domain of integration is $-\infty < \xi, \eta < +\infty.$ Equation~\eqref{eqn-1} is a 2D convolution in the spatial domain and establishes the relationship between the potential field data, $U$ and $V$, recorded at the upward and downward planes, respectively.\\
\indent Assuming that we know the potential field well beyond the lateral extent of all causative bodies, the convolution problem can be equivalently expressed as multiplication in the Fourier domain, as follows:
\begin{equation}
	\label{eqn-4}
	\mathcal{F}[U] = \mathcal{F}[V]  e^{-k \Delta z},
\end{equation}
 where $\mathcal{F}$ denotes the Fourier operator, and $k$ denotes the radial wavenumber, which is defined as $k = \sqrt{k_x^2 + k_y^2}$, with $k_x$ and $k_y$ denoting the wavenumbers along the $x$ and $y$ directions, respectively, which are used in the upward continuation kernel.\\
\indent For the purpose of numerical computation, the integral equation must be discretized. Approximating the integral as a summation over a finite domain $0 \leq \xi \leq \xi_{\text{max}}, 0 \leq \eta \leq \eta_{\text{max}}$ gives rise to the following expression:
	\begin{equation}
		\label{eqn-2}
		{\bf{U}}_{m, n} = \frac{1}{2 \pi}\sum_{i=1}^{M}\sum_{j=1}^{N} {\bf{L}}_{m,n;i, j}{\bf{V}}_{i,j},
	\end{equation}
 where $M$ and $N$ denote the number of rows and columns, respectively, in the discretized integrand, corresponding to the observation points along the $x$ and $y$ directions, respectively; ${\bf{U}}_{m,n} = U(x_m, y_n, \Delta z)$, ${\bf{V}}_{i,j} = V(\xi_i,\eta_j, 0), \xi_i = \displaystyle\frac{\xi_{max}(i-1)}{M-1}, \eta_j = \displaystyle\frac{\eta_{max}(j-1)}{N-1}$; and 

\begin{equation}
\label{eqn-3}
{\bf{L}}_{m,n;i,j} = \frac{\Delta z}{\left((x_m - \xi_i)^2+ (y_n - \eta_j)^2 + \Delta z^2\right)^{\frac{3}{2}}}.
\end{equation}

\begin{figure}[!t]
	\centering
	\includegraphics[width=4.25in]{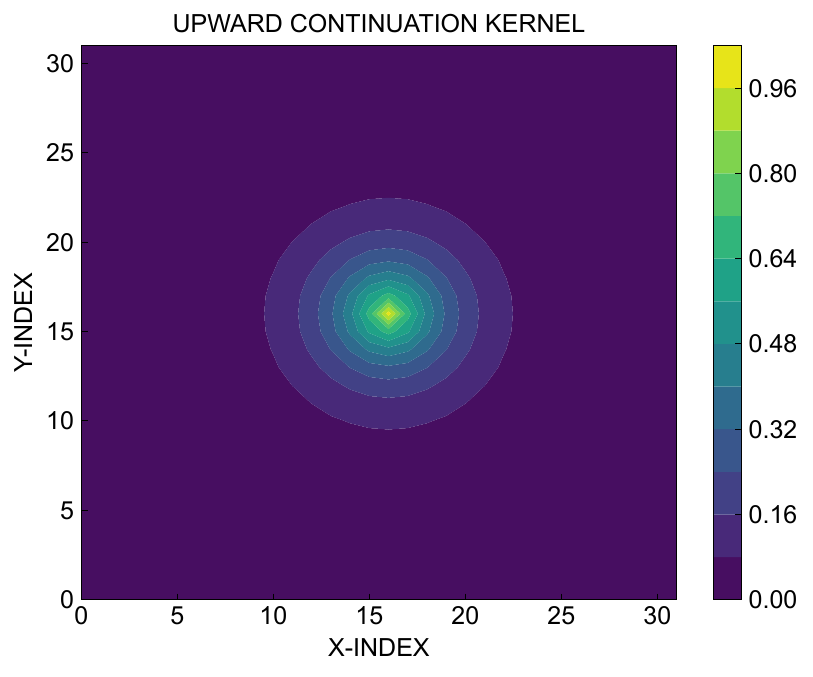}
	\caption{Upward continuation kernel in the Fourier domain for illustrative values of $\Delta z=100$, $\mathrm{d}x=50$.}
	\label{fig2}
\end{figure}

\indent \subsection{Upward Continuation Kernel}
\indent Let $\mathrm{d}x$ and $\mathrm{d}y$ denote the sampling intervals along the $x$ and $y$ directions, respectively. Assuming that the grid has the same even number of points and that the spacing is equal along the $x$ and $y$ directions, that is, $M = N$ and $\mathrm{d}x = \mathrm{d}y$. Let $\bf{a}$ and $\bf{b}$ denote the vectors having the integer values in the range $[-\frac{M}{2}, \frac{M}{2}-1]$ and $[-\frac{N}{2}, \frac{N}{2}-1]$, respectively.

% The wavenumbers in the $x$ and $y$ directions are defined as follows:
% \begin{equation}	
% 	\label{eqn-6}
% 	{\bf{k}}_x^{a} = \frac{2 \pi  n_x^{a}}{M \mathrm{d}x}, \quad {\bf{k}}_y^{a} = \frac{2  \pi  n_y^{a}}{N \mathrm{d}y},
% \end{equation}
% respectively, where $\mathrm{d}x$ and $\mathrm{d}y$ denote the sampling intervals along the $x$ and $y$ directions, respectively, and $n_x$ and $n_y$ are the vectors having the values from the set \{$-\frac{M}{2}, -\frac{M}{2} + 1, ..., \frac{M}{2}-1,  \frac{M}{2}$\} and \{$-\frac{N}{2}, -\frac{N}{2} + 1, ..., \frac{N}{2}-1,  \frac{N}{2}$\}, respectively. The corresponding radial wavenumber is given by
% \begin{equation}	
% \label{eqn-7}
% k = \sqrt{\left(\frac{2 \pi n_x}{M  \mathrm{d}x}\right)^2+ \left(\frac{2 \pi n_y}{N  \mathrm{d}y}\right)^2}.
% \end{equation}
% Eq.~\eqref{eqn-7} simplifies to
% \begin{equation}
% \label{eqn-8}
% k = \frac{2 \pi }{M \mathrm{d}x}\sqrt{n_x^2+n_y^2}.
% \end{equation}
\indent Expanding the expression for radial wavenumber $k$ in Eq.~\eqref{eqn-4}, gives the upward continuation operator, which can be expressed element-wise as follows:
\begin{equation}
\label{eqn-9}
{\bf{P}}_{r,l}^{(\Delta z, \mathrm{d}x)} = \text{exp}\left(- \frac{2 \pi }{M}\frac{\Delta z}{\mathrm{d}x}\sqrt{{\bf{a}}_{r}^2+{\bf{b}}_{l}^2}\right),
\end{equation}
 where ${\bf{P}}$ is the upward continuation kernel. Figure~\ref{fig2} illustrates the upward continuation kernel for illustrative values of $\Delta z$ and $\mathrm{d}x$. Equation~\eqref{eqn-9} illustrates the relationship between grid-spacing and continuation height for the upward continuation kernel. 
\indent The problem of calculating ${\bf{V}}$ from ${\bf{U}}$ is a deconvolution problem, which is mathematically an ill-posed problem. This problem is referred to as {\it downward continuation}. The ill-posedness is what makes deriving an exact analytical solution infeasible.

\subsection{Literature Review and Prior Art}
\indent The methods for downward continuation can be broadly categorized as follows:
\begin{enumerate}
\item {\bf Iterative techniques:}
\indent These techniques \cite{xu2006,guo2020} leverage the discretized kernel matrix, as defined in Eq.~\eqref{eqn-3}, to serve as the forward operator. One of the key advantages of iterative techniques is their computational efficiency when applied to large-scale systems. However, Wang et al. \cite{wang23} highlighted significant challenges faced by iterative methods in reconstructing data with greater continuation depths and for high wavenumber components. In contrast, the proposed method provides superior accuracy in reconstructing downward data, particularly with greater continuation depths and for high wavenumber components.

\item {\bf Frequency-domain methods:} Equation~\eqref{eqn-4} represents the upward continuation problem in the Fourier domain. The formula for downward continuation using Tikhonov regularization, as outlined in \cite{wang23, tikhonov}, is given as follows:
\begin{equation}
\label{eqn-10}
\mathcal{F}[V] = \frac{e^{\Delta z|k|}}{1 + \mu k^2 e^{\Delta z|k|}} \mathcal{F}[U],
\end{equation}
where $\mu$ is the regularization parameter.
Among the notable Fourier-domain methods \cite{tikhonov,dean1958} that have investigated the downward continuation problem, \cite{tikhonov} has demonstrated strong efficacy. A significant challenge associated with the method is in selecting the optimal regularization parameter \cite{chirinos2024} and reconstructing the edges accurately. In contrast, our method does not require parameter tuning after training the network and provides robust edge reconstruction. 
	
\item {\bf Taylor series-based methods:} The Taylor series-based approach includes techniques such as the integrated second vertical derivative (ISVD) \cite{Fedi2002}, and the truncated Taylor-series iterative downward continuation \cite{zhang}. A significant challenge inherent to these methods is the selection of an appropriate regularization parameter \cite{chirinos2024}, such as determining the optimal number of Taylor-series iterations. Moreover, these techniques encounter limitations in accurately reconstructing high-wavenumber components. In contrast, the proposed method provides a robust reconstruction of downward data.

\item {\bf Deep learning-based techniques:}
\indent Recent deep learning-based methods to solve the downward continuation problem have employed the U-Net \cite{Ye2022, Li2023}. By utilizing training data to learn the regularization, deep learning approaches provide a stronger prior compared to the ad hoc selections commonly found in traditional regularization-based techniques. As a result, these methods are expected to deliver more accurate reconstructions of downward data. Additionally, deep learning substantially minimizes the time needed during the prediction phase.\\
\indent Once the network has been trained, parameter adjustments are not needed, unlike traditional regularization-based methods, which require calculating parameters for each input. However, one of the challenges associated with the existing deep-learning methods is that the trained network can yield only a single output. Consequently, a trained network can reconstruct the downward continued data at only one depth level. The diffusion network addresses the limitations of the CNN-based approach, offering the same flexibility by eliminating the need for parameter tuning after training. Additionally, it can reconstruct multiple depth levels using a single trained network while delivering more robust results.
\end{enumerate}

\subsection{Our Contribution}
\indent In this paper, we propose the use of the cold-diffusion model to solve the downward continuation problem. To the best of our knowledge, this is the first work to introduce a diffusion model framework for solving the downward-continuation problem. The proposed method can perform stable downward continuation for synthetic test, field data (out-of-distribution data) and under conditions of correlated noise.

\subsection{Organization of the Paper}
\indent Section~\ref{Cold-diffusion Based Downward Continuation} explores the cold-diffusion-based approach and presents multiple methods for downward continuation utilizing this technique. Section~\ref{Simulation Results} evaluates the effectiveness of cold diffusion for downward continuation, offering comparisons with established approaches and highlighting results across varying continuation heights and levels of noise robustness. Section~\ref{Experiments on Field Data} examines the performance of the proposed method on field data, focusing on the influence of the continuation height and time-steps on its overall effectiveness. Finally, Section~\ref{Conclusion} provides a summary of the experimental findings presented in this article. The notations used in this paper are given in Table~\ref{tab-1} for ready reference.

\begin{table}[t]
	\begin{center}
		\caption{A list of notations used in this paper.}
		\label{tab-1}
		\begin{tabular}{ c | c  }
		\hline 
			Notation & Explanation \\
		\hline 
			${\bf{U}}$ & Data observed at a plane away from the source \\
			${\bf{V}}$ &  Data observed at a plane closer to the source \\
            $M$ & Number of observation points in the $x$-direction \\
            $N$ & Number of observation points in the $y$-direction \\
			$\bf{a}$ & vector (lower-case boldface)\\
			$\bf{b}$ & vector (lower-case boldface)\\   
			$\mathcal{O}$ & Neural network \\
            $\theta$ & Neural network parameter \\
            $t$ & Step-size \\
			$k$ & Radial wavenumber \\
			$\bf{P}$ &  Upward continuation operator \\
			$\Delta z$ & Continuation height \\
            $H$ & Maximum continuation height \\
            $\mathcal{D}$ & Degradation operator\\
			$\mu$ & Regularization parameter\\
			$\mathrm{d}x $ &Sampling step in the $x$-direction\\
			 $\mathrm{d}y$ &Sampling step in the $y$-direction\\
             $\chi$ & Training dataset\\
 		\hline
		\end{tabular}
	\end{center}
\end{table}

% \section{Theory}
\begin{figure*}[htp]
	\centering
	\begin{tikzpicture}[scale=0.7,
	node distance=1cm,
	box/.style={rectangle, draw, minimum width=0.8cm, minimum height=0.8cm, align=center, font=\small},
	arrow/.style={->, thick},
	timeemb/.style={rectangle, draw, fill=blue!20, minimum width=1.5cm, minimum height=0.5cm, align=center, font=\small},
	convblock/.style={rectangle, draw, fill=green!20, minimum width=0.25cm, minimum height=1.2cm, align=center, font=\small},
	sample/.style={rectangle, draw, fill=cyan!20, minimum width=0.25cm, 
		minimum height=1.2cm, align=center, font=\small},
	upsample/.style={rectangle, draw, fill=gray!20, minimum width=0.25cm, 
		minimum height=1.2cm, align=center, font=\small},
	concat/.style={rectangle, draw, fill=teal!20, minimum width=0.25cm, minimum height=1.2cm, align=center, font=\small},
	demons1/.style={rectangle, draw, fill=teal!20, minimum width=1cm, minimum height=0.5cm, align=center, font=\small},
	demons2/.style={rectangle, draw, fill=green!20, minimum width=1cm, minimum height=0.5cm, align=center, font=\small},
	demons3/.style={rectangle, draw, fill=cyan!20, minimum width=1cm, minimum height=0.5cm, align=center, font=\small},
	demons4/.style={rectangle, draw, fill=gray!20, minimum width=1cm, minimum height=0.5cm, align=center, font=\small},
	demons5/.style={rectangle, draw, fill=red!20, minimum width=1cm, minimum height=0.5cm, align=center, font=\small},
	demons6/.style={rectangle, draw, minimum width=1cm, minimum height=0.5cm, align=center, font=\small},
	attention/.style={rectangle, draw, fill=red!20, minimum width=0.25cm, minimum height=1.2cm, align=center, font=\small}
	]

	% Time embedding path
	\node[timeemb] (time) at (0,-2) {TIME INPUT};
	\node[timeemb] (sinemb) at (0,-4) {Sinusoidal \\Positional \\Embedding};
	\node[timeemb] (timemlp1) at (0,-6) {LINEAR};
	\node[timeemb] (timemlp2) at (0,-6.75) {GELU};
	\node[timeemb] (timemlp3) at (0,-7.5) {LINEAR};
	
	\draw[arrow] (time) -- (sinemb);
	\draw[arrow] (sinemb) -- (timemlp1);
	
	% Main U-Net path
	\node[box] (input) at (2,3) {INPUT};
	
	% Downsampling path
	\node[convblock] (down1) at (2,0) {};
	\node[convblock] (down11) at (2.5,0) {};
	\node[attention] (att1) at (3,0) {};
	\node[sample] (down1-sample) at (3.5,0) {};
	
	\node[convblock] (down2) at (3.5,-3.5) {};
	\node[convblock] (down21) at (4,-3.5) {};
	\node[attention] (att2) at (4.5,-3.5) {};
	\node[sample] (down2-sample) at (5,-3.5) {};
	
	\node[convblock] (down3) at (5,-7) {};
	\node[convblock] (down31) at (5.5,-7) {};
	\node[attention] (att3) at (6,-7) {};
	\node[sample] (down3-sample) at (6.5,-7) {};
	
	\node[convblock] (down4) at (6.5,-10.5) {};
	\node[convblock] (down41) at (7,-10.5) {};
	\node[attention] (att4) at (7.5,-10.5) {};
	
	% Middle blocks
	\node[convblock] (mid1) at (11,-10.5) {};
	\node[attention] (midatt) at (11.5,-10.5) {};
	\node[convblock] (mid2) at (12,-10.5) {};
	
	% Upsampling path
	\node[concat](cat1) at (14, -7) {};
	\node[convblock] (up3) at (14.5,-7) {};
	\node[convblock] (up31) at (15,-7) {};
	\node[attention] (upatt3) at (15.5,-7) {};
	\node[upsample] (up3-sample) at (16,-7) {};

	\node[concat](cat2) at (16, -3.5) {};
	\node[convblock] (up2) at (16.5,-3.5) {};
	\node[convblock] (up21) at (17,-3.5) {};
	\node[attention] (upatt2) at (17.5,-3.5) {};
	\node[upsample] (up2-sample) at (18,-3.5) {};
	
	\node[concat](cat3) at (18, 0) {};
	\node[convblock] (up1) at (18.5,0) {};
	\node[convblock] (up11) at (19,0) {};
	\node[attention] (upatt1) at (19.5,0) {};
	\node[upsample] (up1-sample) at (20,0) {};
	
	% final convolution path 
	\node[convblock] (up0) at (20,3) {};
	\node[box] (conv) at (21.5,3) {Conv2D};
	\node[box] (output) at (24,3) {OUTPUT};
	
	%output
	\node[demons1] (d1) at (21,-6) {};
	\node[demons6] (outd1) at (23.5,-6) {CONCAT-\\ENATION};
	\node[demons2] (d2) at (21,-7) {};
	\node[demons6] (outd2) at (23.5,-7) {ConvNext};
	\node[demons3] (d3) at (21,-8.5) {};
	\node[demons6] (outd3) at (23.5,-8.5) {DOWN-\\SAMPLING};
	\node[demons4] (d4) at (21,-10) {};
	\node[demons6] (outd4) at (23.5,-10) {UP-\\SAMPLING};
	\node[demons5] (d5) at (21,-11.5) {};
	\node[demons6] (outd5) at (23.5,-11.5) {LINEAR \\ATTENTION};
	
	% connect its path
	\draw[arrow] (d1) -- (outd1);
	\draw[arrow] (d2) -- (outd2);
	\draw[arrow] (d3) -- (outd3);
	\draw[arrow] (d4) -- (outd4);
	\draw[arrow] (d5) -- (outd5);
	
	% Connect main path
	\draw[arrow] (input) -- (down1);
	\draw[arrow] (down1) -- (down11);
	\draw[arrow] (down11) -- (att1);
	\draw[arrow] (att1) -- (down1-sample);
	\draw[arrow] (down1-sample) -- (down2);
	\draw[arrow] (down2) -- (down21);
	\draw[arrow] (down21) -- (att2);
	\draw[arrow] (att2) -- (down2-sample);
	\draw[arrow] (down2-sample) -- (down3);
	\draw[arrow] (down3) -- (down31);
	\draw[arrow] (down31) -- (att3);
	\draw[arrow] (att3) -- (down3-sample);
	\draw[arrow] (down3-sample) -- (down4);
	\draw[arrow] (down4) -- (down41);
	\draw[arrow] (down41) -- (att4);
	
	% Middle connections
	\draw[arrow] (att4) -- (mid1);
	\draw[arrow] (mid1) -- (midatt);
	\draw[arrow] (midatt) -- (mid2);
	
	% Upsampling connections
	\draw[arrow] (mid2) -| ($(mid2)+(1,0)$) -| (cat1);
	\draw[arrow] (cat1) -- (up3);
	\draw[arrow] (up3) -- (up31);
	\draw[arrow] (up31) -- (upatt3);
	\draw[arrow] (upatt3) -- (up3-sample);
	\draw[arrow] (up3-sample) -- (cat2);
	\draw[arrow] (cat2) -- (up2);
	\draw[arrow] (up2) -- (up21);
	\draw[arrow] (up21) -- (upatt2);
	\draw[arrow] (upatt2) -- (up2-sample);
	\draw[arrow] (up2-sample) -- (cat3);
	\draw[arrow] (cat3) -- (up1);
	\draw[arrow] (up1) -- (up11);
	\draw[arrow] (up11) -- (upatt1);
	\draw[arrow] (upatt1) -- (up1-sample);
	
	% Final connections
	% \draw[arrow] (up1_sample) -- (up1_sample);
	
	\draw[arrow] (up1-sample) -- (up0);
	\draw[arrow] (up0) -- (conv);
	\draw[arrow] (conv) -- (output);
	
	% Skip connections
		\draw[arrow, dashed] (down1-sample) -- (cat3);
		\draw[arrow, dashed] (down2-sample) -- (cat2);
		\draw[arrow, dashed] (down3-sample) -- (cat1);
%	\draw[arrow] (down1-sample) -| ($(cat3)-(0.2, 0.2)$) -| (cat3);
%	\draw[arrow, dashed] (down2-sample) -| ($(up2)+(0,0.3)$);
%	\draw[arrow, dashed] (down3-sample) -| ($(up3)+(0,0.3)$);

	\coordinate (timemlp-out) at ($(timemlp3)+(0,-0.75)$);
	\draw[arrow, blue] (timemlp3) -- (timemlp-out);
	% Time embedding connections
	\draw[arrow, blue] (timemlp-out) -| ($(timemlp3)+(0,-1)$) -| (down1);
	\draw[arrow, blue] (timemlp-out) -| ($(timemlp3)+(0,-1)$) -| (down11);
	\draw[arrow, blue] (timemlp-out) -| ($(timemlp3)+(0,-2)$) -| (down2);
	\draw[arrow, blue] (timemlp-out) -| ($(timemlp3)+(0,-2)$) -| (down21);
	\draw[arrow, blue] (timemlp-out) -| ($(timemlp3)+(0,-3)$) -| (down3);
	\draw[arrow, blue] (timemlp-out) -| ($(timemlp3)+(0,-3)$) -| (down31);
	\draw[arrow, blue] (timemlp-out) -| ($(timemlp3)+(0,-5)$) -| (down4);
	\draw[arrow, blue] (timemlp-out) -| ($(timemlp3)+(0,-5)$) -| (down41);
	\draw[arrow, blue] (timemlp-out) -| ($(timemlp3)+(0,-5)$) -| (mid1);
	\draw[arrow, blue] (timemlp-out) -| ($(timemlp3)+(0,-5)$) -| (mid2);
	\draw[arrow, blue] (timemlp-out) -| ($(timemlp3)+(0,-5.25)$) -| (up3);
	\draw[arrow, blue] (timemlp-out) -| ($(timemlp3)+(0,-5.25)$) -| (up31);
	\draw[arrow, blue] (timemlp-out) -| ($(timemlp3)+(0,-5.5)$) -| (up2);
	\draw[arrow, blue] (timemlp-out) -| ($(timemlp3)+(0,-5.5)$) -| (up21);
	\draw[arrow, blue] (timemlp-out) -| ($(timemlp3)+(0,-5.75)$) -| (up1);
	\draw[arrow, blue] (timemlp-out) -| ($(timemlp3)+(0,-5.75)$) -| (up11);
	\draw[arrow, blue] (timemlp-out) -| ($(timemlp3)+(0,-6)$) -| (up0);
\end{tikzpicture}
\caption{U-Net architecture used for cold-diffusion.}
\label{fig3}
\end{figure*}
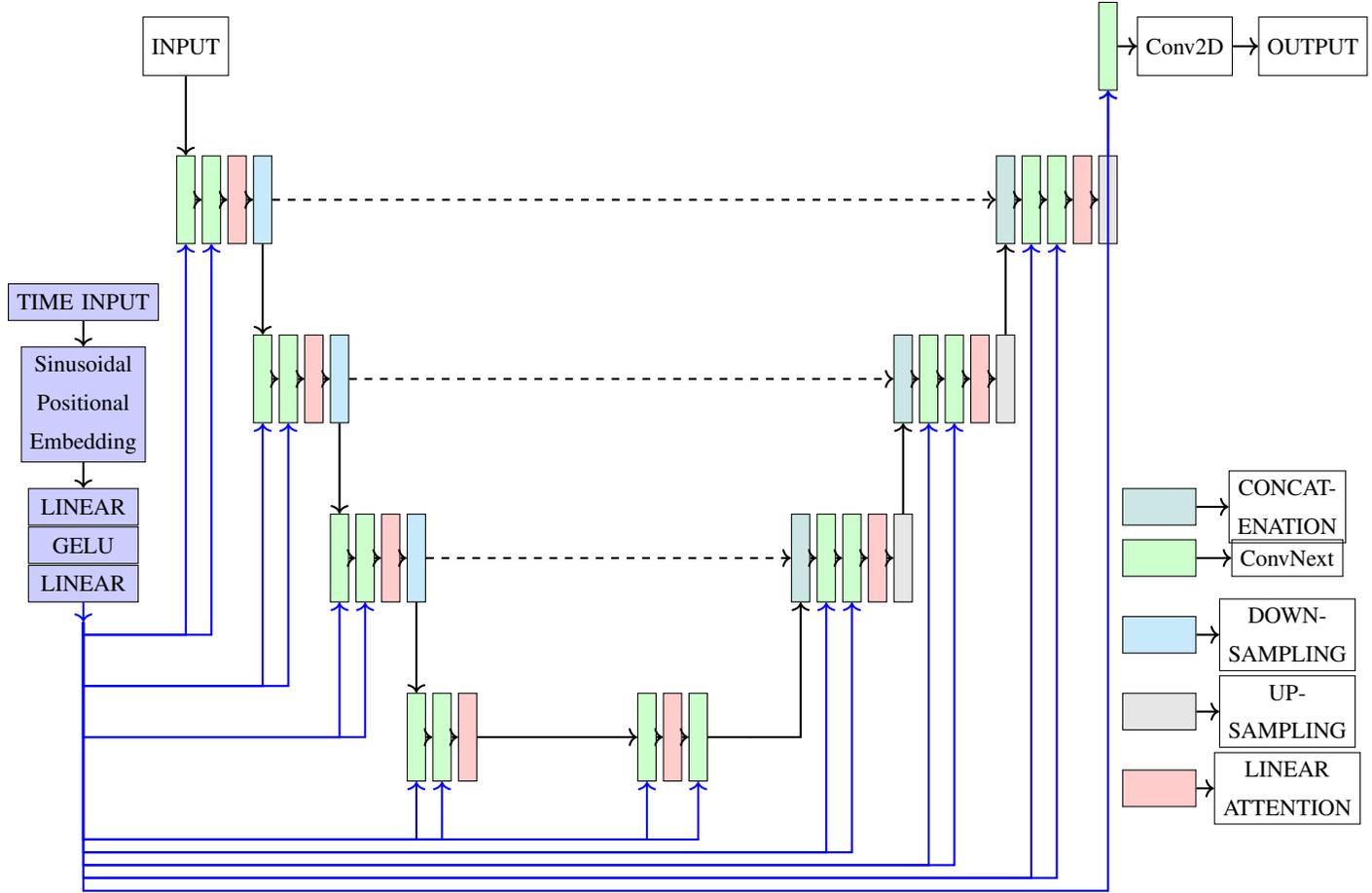

\section{Cold-diffusion Based Downward Continuation}
\label{Cold-diffusion Based Downward Continuation}
\begin{figure}[t]
	\centering
	\includegraphics[width=4.25in]{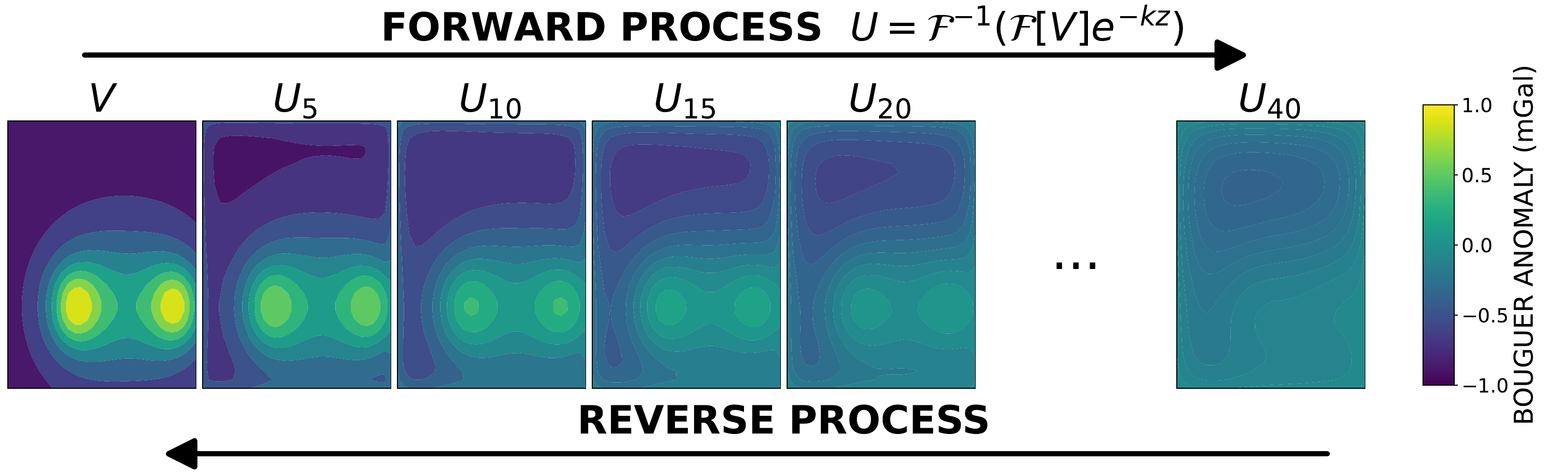}
	\caption{Schematic representation of diffusion steps generated with the upward continuation kernel.}
	\label{fig4}
\end{figure}

\indent Diffusion models, originally introduced as the Denoising Diffusion Probabilistic Model (DDPM) \cite{Ho2020}, involve progressively adding Gaussian noise to the input in a bid to learn its distribution. Diffusion models operate in two phases: the forward phase, where noise is gradually added to the image; and the reverse phase, which progressively removes the noise. The primary goal of diffusion models is to generate new samples from the learned distribution. Cold diffusion \cite{colddif}, introduced recently, extends the application of diffusion models to address inverse problems and invert arbitrary transformations. We employ the cold-diffusion framework for performing downward continuation of gravity data. Figure~\ref{fig3} illustrates the U-Net network utilized as the backbone in implementing the cold-diffusion network.\\
Consider dividing the continuation height $\Delta z$ into $h+1$ steps, with $t$ as the step-size such that $h \times t = \Delta z$. Let ${\bf{U}}_h$ represent the upward continued data at a continuation height of $h$ steps, ${\bf{V}}$ denote the data at the downward plane, and $\mathcal{D}$ be the degradation operator, defined as $\mathcal{D}({\bf{V}}, h) = {\bf{U}}_h$. In practice, the operator $\mathcal{D}$ is implemented as follows:
\begin{equation}
\label{eqn-12}
    \mathcal{D}({\bf{V}}, h) = \mathcal{F}^{-1}[{{\bf{P}}^{(h \times t,~\mathrm{d}x)}} \times \mathcal{F}[{\bf{V}}]],
\end{equation}
where $\mathcal{F}^{-1}$ denotes the inverse Fourier operator, and $P$ represents the upward continuation kernel defined by Eq.~\eqref{eqn-9}.\\
\indent Let $\mathcal{O}$ denote the neural network that inverts the degradation operator $\mathcal{D}$. The network $\mathcal{O}$ is implemented with $\theta$ denoting its parameters. Let $\mathcal{X}$ denote the set of images in the training dataset, and $H$ be the maximum continuation height, and $\gamma$ be the number of steps corresponding to $H$, such that $\gamma\,t = H$. The loss function optimization for training the cold diffusion network is given by:
\begin{equation}
    \label{eqn-13}
    \min_{\theta} \mathop{\mathbb{E}}_{h \in S}\mathop{\mathbb{E}}_{V \in \mathcal{X}}||\mathcal{O}_\theta(\mathcal{D}({\bf{V}},h),h) - {\bf{V}}||_1,
\end{equation}
 where $h$ is sampled from a uniform distribution over the set $S$, where $S = \{0, 1, \ldots, \gamma-1, \gamma\}$, $\mathbb{E}$ denotes the expectation operator, and $\|\cdot\|_1$ denotes the $l_1$ norm. Figure~\ref{fig4} illustrates the upward and downward continuation as the forward and reverse processes of the diffusion model, respectively.
\indent Next, we introduce three approaches for applying the trained neural network to perform downward continuation: (i) Direct reconstruction; (ii) Recursive downward continuation; and (iii) Stabilized downward continuation. In Section~\ref{Simulation Results}, we evaluate the performance of these three methods. In all the three methods, $\alpha_\gamma$ denotes the observed data, which serves as the input, and $\alpha_0$ denotes the output of the neural network, which should be the data at the surface. In the second and third methods, ${\bf{V}}_s$ represents the reconstruction of downward data through the network $\mathcal{O}$ for input at the $s^{th}$ height step.

\subsection{Direct Reconstruction}
\indent In this method, we supply the observed data along with the corresponding continuation height towards the downward plane, and the trained neural network $\mathcal{O}$ produces an approximation of the data closer to the source:
\begin{equation}
\label{eqn-14}
    \mathcal{O}(\alpha_\gamma, \gamma) = {\alpha}_0,
\end{equation}
where ${\alpha}_0$ is an approximation of ${\bf{V}}$.
 
\subsection{Recursive Downward Continuation}
\indent This method iteratively reconstructs the data using the trained model $\mathcal{O}$. This method is inspired by the naive sampling procedure outlined in Algorithm~\ref{alg1} \cite{Ho2020,Song2021}. This algorithm inputs the observed data and the step corresponding to the continuation height and finally returns the approximation of the data in the downward plane.

\subsection{Stabilized Downward Continuation}
\indent Algorithm~\ref{alg2} is similar to the transformation-agnostic cold sampling (TACoS) introduced by Bansal et al. \cite{colddif}. This algorithm processes the observed data along with the continuation height and performs iterative reconstruction by adding the difference between the data at the upward plane and the upward continued reconstructed data at each step. The difference between the blurred data and the upward continued reconstructed data may represent edge information or high wavenumber features of the gravity data lost during the upward continuation. Incorporating these high wavenumber features at each step retains sharp features of the gravity data. Thus, this method is expected to outperform the recursive DC algorithm.

\section{Simulation Results}
\label{Simulation Results}
\indent \subsection{Model Training}
\label{Model Training}
\indent We generated 22,000 training samples of the gravity data, which constitutes $\chi$, our training dataset, and their corresponding models using the random walk-based methodology introduced by Huang et al. \cite{huangetal}. The cold-diffusion network ($\mathcal{O}$) was employed for downward continuation and trained for 100,000 steps (approximately 146 epochs) using the loss function given by Eq.~\eqref{eqn-13} and Adam optimizer \cite{kingma2015}. The training was performed on an NVIDIA RTX 3090 GPU with 24 GB memory. 
\begin{algorithm}[t]
	\caption{Recursive downward continuation.}\label{alg:alg1}
	\begin{algorithmic}
		\STATE 
		\STATE {\textbf{Input: }}Observed data $\alpha_\gamma$
		\STATE {\textbf{for}}$\hspace{0.25cm}s = \gamma, \gamma-1,\ldots, 1$ \textbf {do}
		\STATE \hspace{0.5cm}${\bf{V}}_s \leftarrow \mathcal{O}(\alpha_s, s)$ \hspace{0.9cm}{\textbf{(reconstruction step)}}
		\STATE \hspace{0.5cm}$\alpha_{s-1} = \mathcal{D}({\bf{V}}_s, s-1)$ {\textbf{(degradation step)}}
		\STATE {\textbf{End for}}
		\STATE {\textbf{Return: }}$\alpha_0$
	\end{algorithmic}
	\label{alg1}
\end{algorithm}

\begin{algorithm}[t]
	\caption{Stabilized downward continuation.}\label{alg:alg2}
	\begin{algorithmic}
		\STATE 
		\STATE {\textbf{Input: }}Observed data $\alpha_\gamma$
		\STATE {\textbf{for}}$\hspace{0.25cm}s = \gamma, \gamma-1,\ldots, 1$ \textbf {do}
		\STATE \hspace{0.5cm}${\bf{V}}_s \leftarrow \mathcal{O}(\alpha_s, s)$\hspace{1.5cm}{\textbf{(reconstruction step)}}
		\STATE \hspace{0.5cm}$\alpha_{s-1} = \alpha_s -  \mathcal{D}({\bf{V}}_s, s) + \mathcal{D}({\bf{V}}_s, s-1)$\\\hspace{0.5cm}{\textbf{(degradation step)}}
		\STATE {\textbf{End for}}
		\STATE {\textbf{Return: }}$\alpha_0$
	\end{algorithmic}
	\label{alg2}
\end{algorithm}

\subsection{Quantitative Metrics}
\indent  For quantitative evaluation of the performance of the cold-diffusion network, we employ the peak signal-to-noise ratio (PSNR) and structural similarity index measure (SSIM) \cite{Bovik2004}. Additionally, we include other metrics such as the relative error (RE) and root mean-square error (RMSE). The mathematical expressions for these metrics are provided in the appendix for ready reference.

\subsection{Comparison of DC methods}
\indent In this subsection, we first present the workflow of the various downward continuation methods used for benchmarking the performance with respect to the cold-diffusion approach. Subsequently, we will outline the setup for generating test data and discuss the quantitative performance of the methods using the test data. Specifically, we employed the following techniques: (i) Tikhonov regularized downward continuation (TRDC) \cite{tikhonov}; (ii) Truncated Taylor series iterative downward continuation (TTSIDC) \cite{zhang}; (iii) DC-Net \cite{Li2023}; and (iv) D-U-Net \cite{Ye2022}.\\
\indent Wang et al. \cite{wang23} demonstrated through experiments that the TRDC \cite{tikhonov} and TTSIDC \cite{zhang} outperform conventional downward continuation methods, which is why we select them for performance comparison. In our experiments, the C-norm \cite{tikhonov} was employed to optimize the regularization parameter by minimizing the RMSE between the downward data and input data, thereby determining the optimal solutions for TRDC and TTSIDC, leading to the implementation of oracle versions of these methods in our experiments.\\
\indent In the realm of deep learning-based downward continuation networks, we have utilized both DC-Net \cite{Li2023} and D-U-Net \cite{Ye2022} in our implementation. For the D-U-Net model, we employed the single-input configuration. Additionally, we also implemented the U-Net, which serves as the backbone of the cold-diffusion network. All three U-Net-type architectures were trained using the same training dataset $\chi$ for 150 epochs.

\begin{figure}[!t]
	\centering
	\includegraphics[width=3.25in]{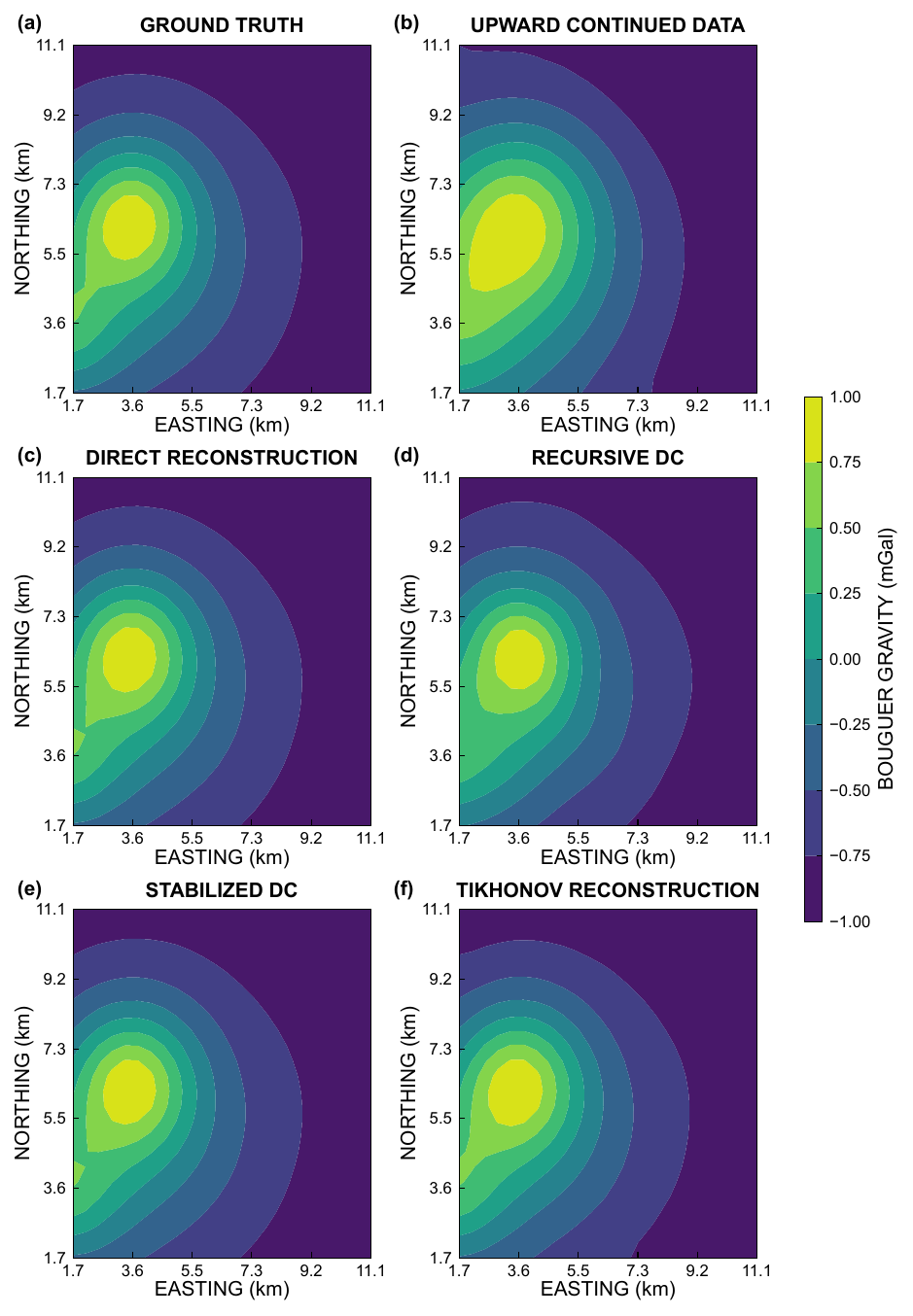}
	\caption{Qualitative evaluation of different techniques for downward continuation. The images have been normalized to the range [$-1, 1$] for the purpose of visualization.}
	\label{fig5}
\end{figure}

\indent Table~\ref{tab-2} presents a quantitative comparison of the downward continuation performance among TRDC, three U-Net-based architectures, and cold-diffusion results, including direct reconstruction, recursive DC, and stabilized DC. The synthetic test dataset comprises 10 geologically plausible models, generated with the subsurface discretization of $50\times50\times50$ and an observation plane of $32\times32$, with a grid spacing of $50\,$m. We averaged the quantitative performance across 10 samples from the test dataset, each continued upward to a height of $100\,$m for all the aforementioned methods. The experimental results are presented in Table~\ref{tab-2}, which show that the cold-diffusion network, with stabilized DC and direct reconstruction, surpasses the performance of the U-Net-based DC technique. The reconstruction achieved by stabilized DC shows significant improvement over that obtained through recursive DC. Figure~\ref{fig5} depicts the qualitative results of downward continuation using the TRDC and the cold-diffusion approaches. To facilitate qualitative comparisons (Figures~\ref{fig5}, \ref{fig9}, and \ref{fig-qualitative-correlated-noise}) and quantitative comparisons (Tables~\ref{tab-2}, \ref{tab-3}, and Figure~\ref{fig-quantitaive-correlated-noise}) of our proposed approach with other techniques, we utilized $24\times24$ central image patch to account for the challenges that conventional techniques typically encounter at the image boundaries. The reconstructed images for all the techniques listed in Table~\ref{tab-2} are available in  Supplementary Material.

\begin{table*}[t]
	\begin{center}
		\caption{Comparison of different downward continuation approaches on synthetic data. The performance metrics and corresponding standard deviation are indicated. The best performance is highlighted in bold. }
		\label{tab-2}
		\begin{tabular}{| c | c | c | c | c | }
			\hline
			Approach& PSNR (dB)$\uparrow$ & SSIM $\uparrow$& RE $\downarrow$& RMSE $\downarrow$ \\
			%			of filter & $e_m$ &   $b_{ij}$ & 1 & 1 & 1\\
			\hline 
			TTSIDC \cite{zhang} & $32.4564 \pm 3.0311$ & $0.9611 \pm 0.0281$ & $0.0738 \pm 0.0191$ & $0.0498 ± 0.0191$ \\
			\hline
			TRDC \cite{tikhonov}&  $36.5660 \pm 2.3506$ & $0.9772 \pm 0.0085$ &  $0.0452 \pm 0.0080$ & $0.0302 \pm 0.0095$ \\
			\hline
			D-U-Net \cite{Ye2022}& $47.1495 \pm 0.9401$ & $0.9975 \pm 0.0011$& $0.0135 \pm 0.0029$ & $0.0086 \pm 0.0009$ \\
			\hline 
			DC-Net \cite{Li2023}& $50.7815 \pm 3.7485$ & $0.9993 \pm 0.0004$ & $0.0095 \pm 0.0048$ & $0.0064 \pm 0.0039$ \\ 
			\hline
			U-Net \cite{colddif}& $52.3242  \pm 5.2551$ & $0.9993 \pm 0.0003$ & $0.0087 \pm 0.0064$ & $0.0060 \pm 0.0056$ \\
			\hline
			CD (Direct) -- Ours& $56.5956 \pm 4.6913$ & $0.9998 \pm 0.0003$ &  $0.0050 \pm 0.0032$ & $0.0035 \pm 0.0027$ \\
			\hline
			CD (Stabilized DC) -- Ours&  $\bf{61.2089 \pm 7.1674}$ & $\bf{0.9999 \pm 0.0002}$ & $\bf{0.0037 \pm 0.0042}$ & $\bf{0.0027 \pm 0.0034}$ \\
			\hline
			CD (Recursive DC) -- Ours& $32.8684 \pm 2.8821$ & $0.9518 \pm 0.0403$ & $0.0742 \pm 0.0298$ & $0.0466 \pm 0.0152$ \\
			\hline
		\end{tabular}
		% \caption*{The best performances are highlighted in bold.}
	\end{center}
\end{table*}

\subsection{Performance for Different Continuation Heights}
\begin{figure}[!t]
	\centering
	\includegraphics[width=4.25in]{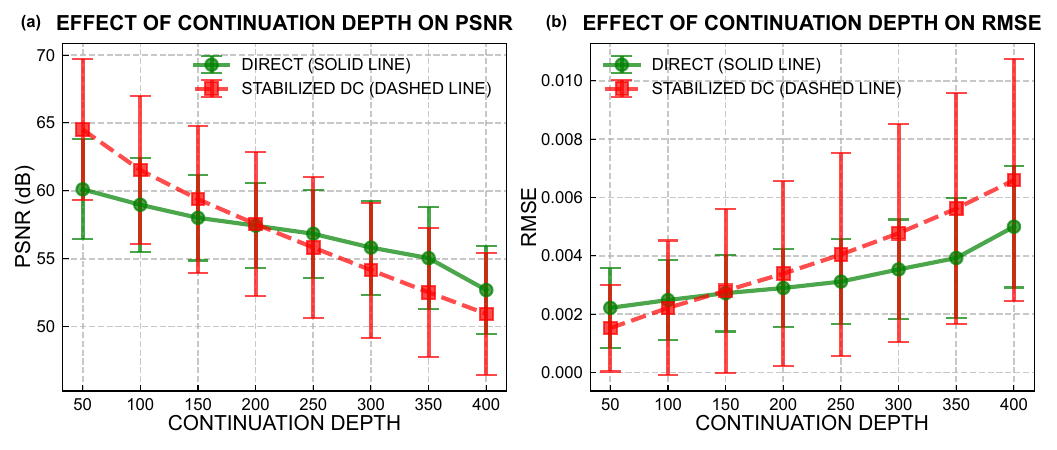}
	\caption{Quantitative performance of cold-diffusion technique on synthetic test data across different continuation heights.}
	\label{fig6}
\end{figure}

\indent We now describe the experimental setup used to assess the performance of model $\mathcal{O}$ for DC at various continuation heights, followed by a discussion of the results. The cold-diffusion model ($\mathcal{O}$) was trained using the training dataset $\chi$, where observed data are at a continuation height of $400\,$m towards the downward plane. This continuation height is divided into 40 height steps with a step-size of $10\,$m. To evaluate the performance of the model, we tested it on 10 different synthetic datasets, averaging the results to alleviate bias toward any particular dataset. Figures~\ref{fig6}a and \ref{fig6}b demonstrate that the single-trained model reliably reconstructs data across various continuation heights, with PSNR values exceeding 50 and RMSE values remaining below 0.08 for all heights. In these figures, `direct' and `stabilized DC' correspond to the direct reconstruction and the stabilized DC methods, respectively, of the cold-diffusion approach.

\begin{figure}[!t]
	\centering
	\includegraphics[width=3.25in]{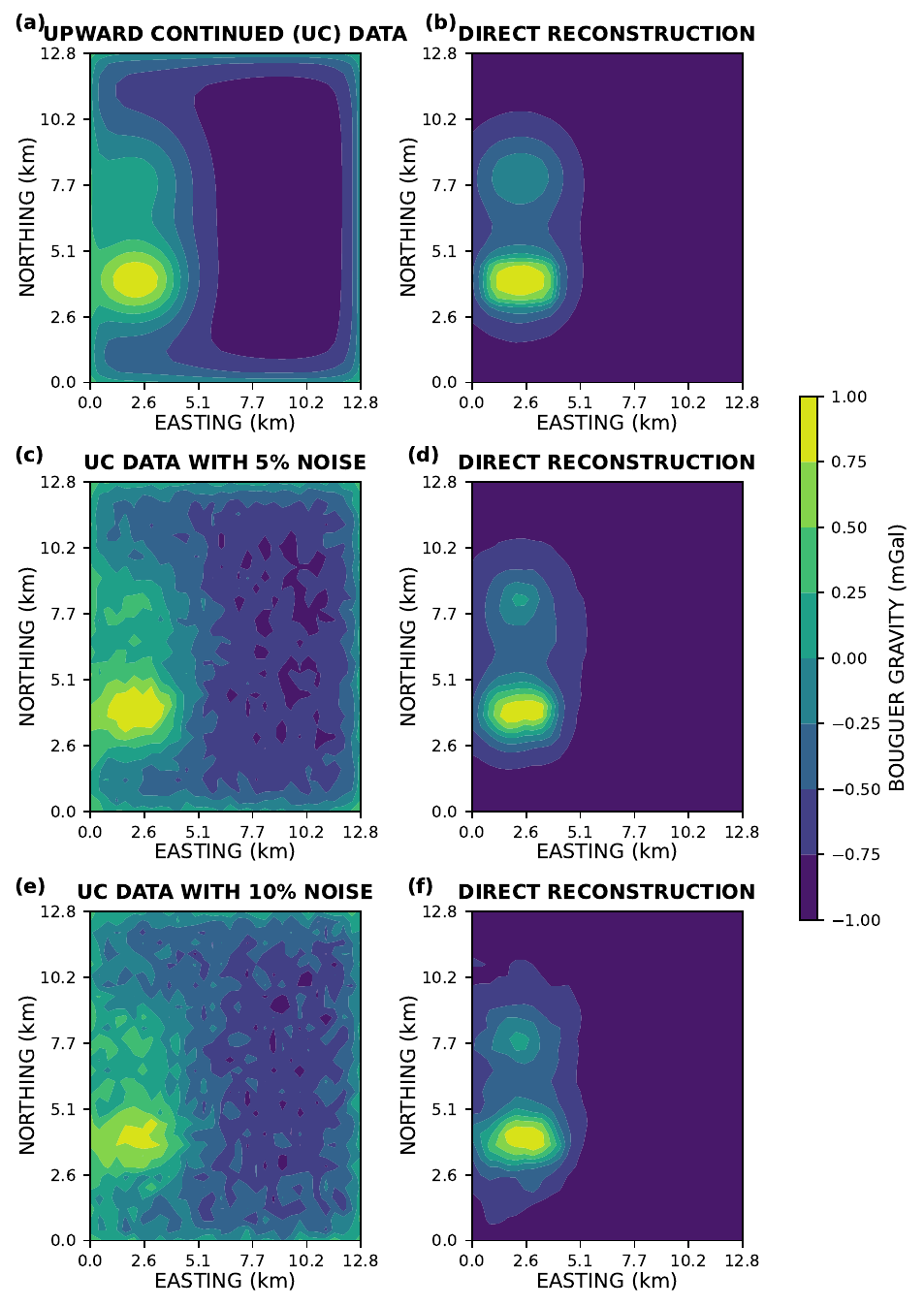}
	\caption{Qualitative performance of cold-diffusion network on the noisy data with different noise levels. The images have been normalized to the range [$-1, 1$] for the purpose of visualization.}
	\label{fig7}
\end{figure}

\begin{figure}[!t]
	\centering
	\includegraphics[width=4.25in]{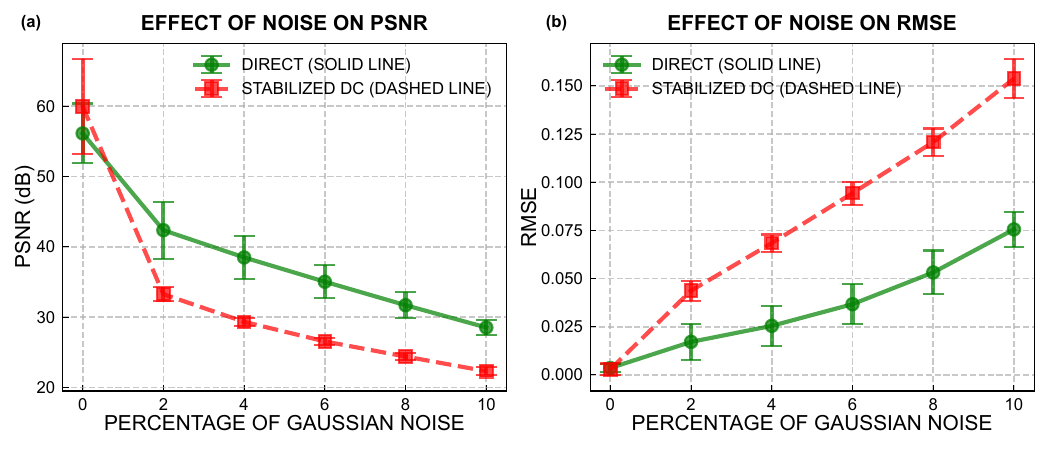}
	\caption{Quantitative evaluation of cold-diffusion network with different noise levels.}
	\label{fig8}
\end{figure}

\subsection{Noise Robustness}
\indent Noise robustness was evaluated by generating training data in accordance with the methodology described in Section~\ref{Model Training}, with noise introduced during training. The standard deviation of the Gaussian noise was determined by multiplying the specified noise percentage by the range of the upward continuation data. During training, we incorporated 2\% Gaussian noise for 10\% of the data, and 4\% Gaussian noise for another 10\% of the data, thereby enabling the model to develop the capability to manage noise effectively. Figure~\ref{fig7} depicts the qualitative performance of the cold-diffusion network on noisy data at varying noise levels. Figures~\ref{fig8}a and \ref{fig8}b demonstrate that the model successfully maintained an RMSE below 0.08 even when subjected to 10\% Gaussian noise. Despite being trained with 2\% and 4\% Gaussian noise, the reconstruction capability of the model, even in the presence of intense noise, demonstrates the noise robustness of the model. Additionally, the direct reconstruction results in superior noise robustness capability compared with stabilized DC. 

\section{Experiments on Field Data}
\label{Experiments on Field Data}
\indent Gravity survey data from the Mangampeta region was used to assess the feasibility of the proposed approach. The Mangampeta Baryte deposit is located in the southern part of the Cuddapah basin in the Cuddapah district, Andhra Pradesh, India. The data was provided by the Geological Survey of India \cite{Ganguli2019}. We applied upward continuation to the ground gravity data as the forward operation to achieve a smoothness similar to that observed in airborne data. The gravity data was gridded with a sampling distance of $400\,$m on both axes. To replicate the smoothing effect observed in airborne measurements collected at an altitude of $100\,$m with a spacing of $50\,$m, Eq.~\eqref{eqn-9} suggests that the parameters $\Delta z$ and $\mathrm{d}x$ should maintain a consistent $\Delta z/\mathrm{d}x$ ratio. Consequently, given the $400\,$m sampling interval, we applied an upward continuation to a height of $800\,$m.

\begin{figure}[!t]
	\centering
	\includegraphics[width=3.25in]{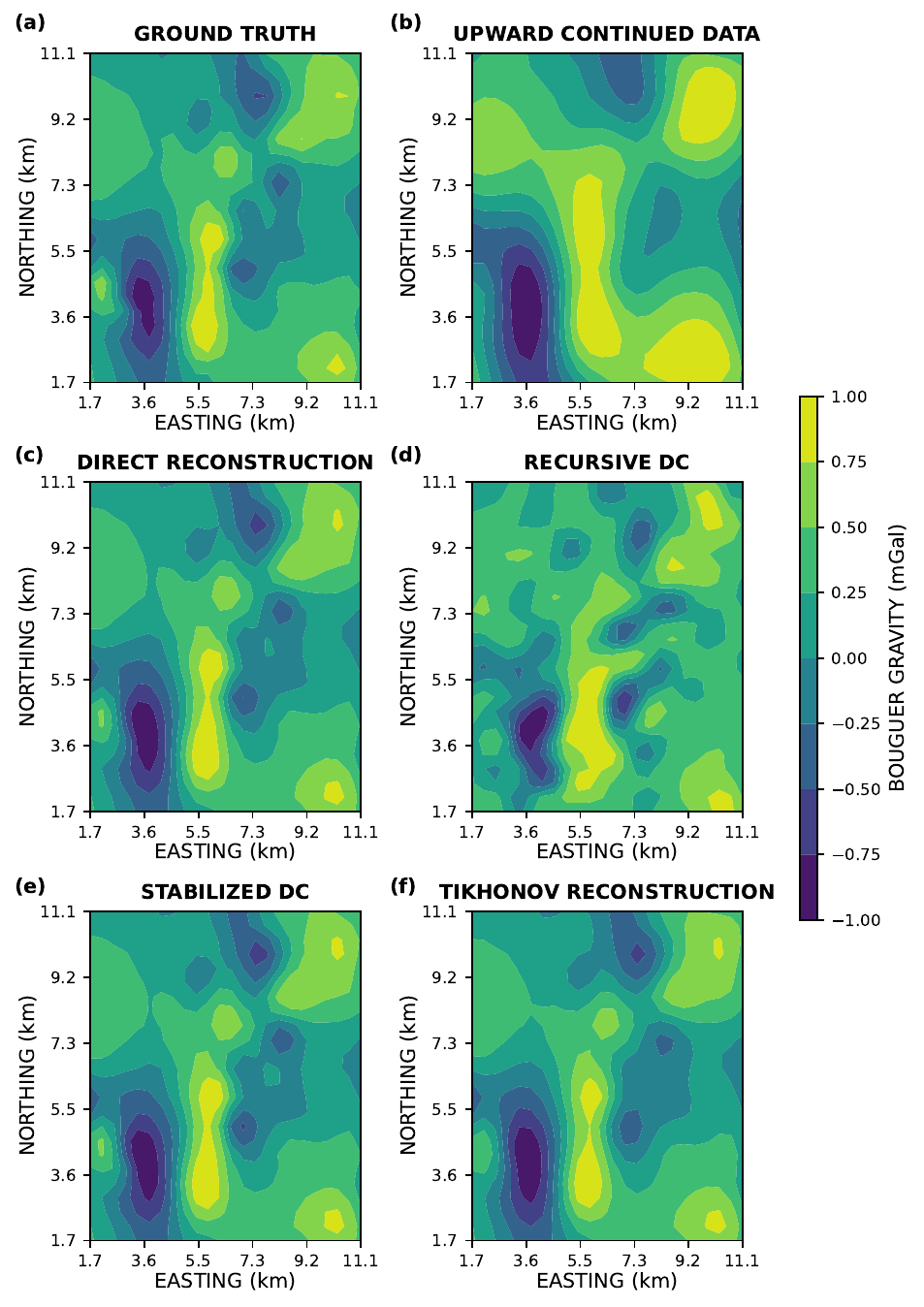}
	\caption{Qualitative assessment of various DC techniques on field data.}
	\label{fig9}
\end{figure}

\subsection{Comparison of Different DC Methods}
\label{comparison-field-data-noiseless}
\indent The training data for the cold-diffusion network was generated using discretized subsurface blocks, each with dimensions of $400 \times 400 \times 200$. The observation plane was configured to dimensions of $32 \times 32$, with a sampling distance of $400\,$m, an upward continuation of $800\,$m, and 40 steps with a step size of $80\,$m. The model $\mathcal{O}$ was trained using the loss function described in Eq.~\eqref{eqn-13} and Adam optimizer \cite{kingma2015}.\\
\indent Table~\ref{tab-3} presents the quantitative performance metrics of various downward continuation methods applied to the field data. It is observed that the proposed cold-diffusion-based method demonstrates superior quantitative performance compared with existing downward continuation methods. Figure~\ref{fig9} shows the qualitative results of downward continuation using the TRDC and the cold-diffusion approach. The reconstruction results for the techniques listed in Table~\ref{tab-3}, along with outcomes on publicly available field gravity data, are provided in Supplementary Material.

\begin{table*}[t]
	\begin{center}
		\caption{Comparison of different downward continuation approaches on field data. The performance metrics and corresponding standard deviation are indicated. The best performance is highlighted in bold.}
    	\label{tab-3}
		\begin{tabular}{| c | c | c | c | c | }
			\hline
			Approach& PSNR (dB)$\uparrow$ & SSIM $\uparrow$& RE $\downarrow$& RMSE $\downarrow$\\
			%			of filter & $e_m$ &   $b_{ij}$ & 1 & 1 & 1\\
			\hline
			TTSIDC \cite{zhang} & $30.1345 \pm 1.6646$ & $0.9489 \pm 0.0210$ & $0.1551 \pm 0.0274$ & $0.0546 \pm 0.0119$\\
			\hline
			TRDC \cite{tikhonov}& $32.4098 \pm 2.1364$ & $0.9611 \pm 0.0202$ & $0.1206 \pm 0.0280$ & $0.0427 ± 0.0124$ \\
			\hline
			D-U-Net \cite{Ye2022}& $32.7571 \pm 2.0261$ & $0.9624 \pm 0.0154$ & $0.1161 \pm 0.0260$ & $0.0410 \pm 0.0107$ \\
			\hline
			DC-Net \cite{Li2023}& $33.0254 \pm 1.5665$ & $0.9683 \pm 0.0129$ & $0.1118 \pm 0.0235$  & $0.0392 \pm 0.0093$ \\ 
			\hline 
			U-Net \cite{colddif}& $31.8232 \pm 2.0487$ & $0.9615 \pm 0.0179$ & $0.1306 \pm 0.0342$ & $0.0459 \pm 0.0132$ \\
			\hline
			CD (Direct) -- Ours& $\bf{36.9401 \pm 2.7436}$ & $\bf{0.9863 \pm 0.0110}$& $\bf{ 0.0737 \pm 0.0244}$ & $\bf{0.0262 \pm 0.0101}$ \\
			\hline
			CD (Stabilized DC) -- Ours& $36.1959 \pm 2.6886$ & $0.9844 \pm 0.0119$ & $ 0.0800 \pm 0.0254$ & $0.0284 \pm 0.0106$ \\
			\hline
			CD (Recursive DC) -- Ours & $19.3599 \pm 2.4070$ & $0.6048 \pm 0.1246$ & $0.5570 \pm 0.1915$ & $0.1919 \pm 0.0610$ \\
			\hline
		\end{tabular}
	\end{center}
\end{table*}

\begin{figure}[!t]
	\centering
	\includegraphics[width=4.25in]{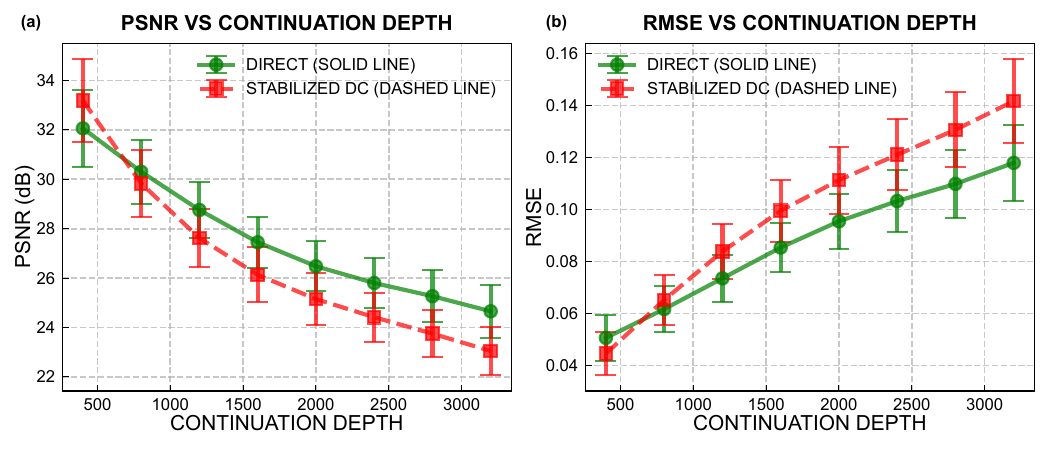}
	\caption{Quantitative performance of a cold-diffusion model on field data across different continuation heights.}
	\label{fig10}
\end{figure}
\subsection{Robustness to Correlated Noise}
\begin{figure}[!t]
\centering
\includegraphics[width=3.25in]{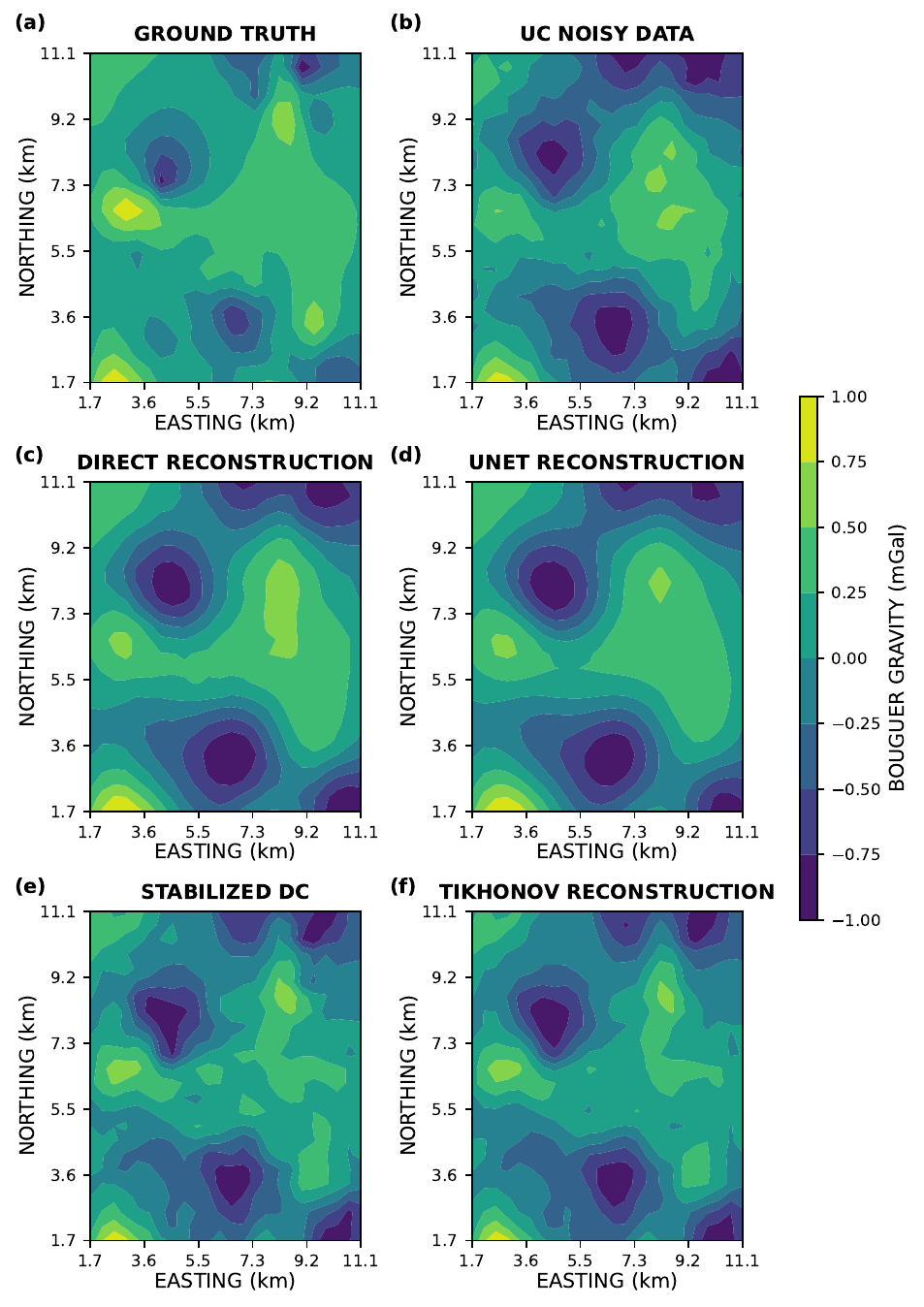}
	\caption{Qualitative assessment of correlated noise on the field data. The Tikhonov reconstruction corresponds to the {\it oracle} scenario where the ground-truth is used to select the regularization parameter.}
\label{fig-qualitative-correlated-noise}
\end{figure}

\begin{figure}[!t]
	\centering
	\includegraphics[width=3.25in]{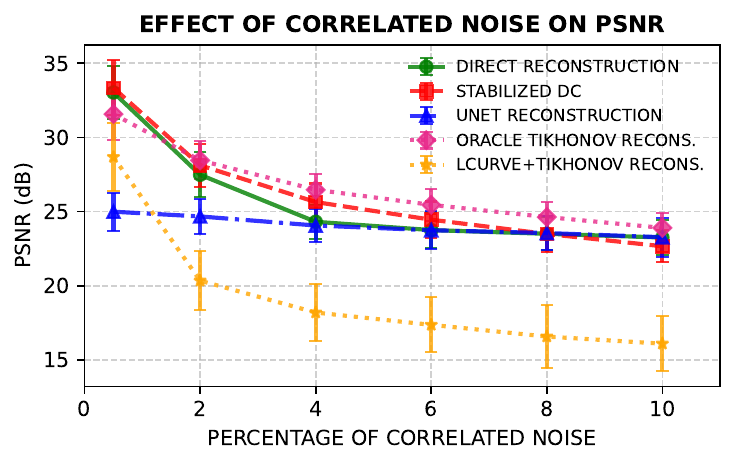}
	\caption{Quantitative performance of a cold-diffusion model on field data across different levels of correlated noise.}
	\label{fig-quantitaive-correlated-noise}
\end{figure}

\indent In practical scenarios, noise affecting measurements is often not white and may be correlated with the data. To evaluate the robustness of our approach under such conditions, we conducted experiments using Gaussian correlated noise with correlation length equal to 0.5. For fine-tuning, we initialized the cold-diffusion model $\mathcal{O}$ and U-Net with the parameters used in Section~\ref{comparison-field-data-noiseless}. Model $\mathcal{O}$ was subsequently trained for 60,000 steps, and the U-Net was fine-tuned over 120 epochs using noisy data. Specifically, 2\% i.i.d. Gaussian noise was added to 10\% of the training data, and 4\% i.i.d. Gaussian noise was added to another 10\%, allowing the model to adapt to varying noise levels. Figure~\ref{fig-qualitative-correlated-noise} illustrates the impact of 5\% correlated noise on various DC techniques applied to field data. As shown in Figure~\ref{fig-quantitaive-correlated-noise}, the performance of the stabilized DC method is better than U-Net and on par with the {\it oracle} Tikhonov reconstruction that assumes knowledge of the ground-truth, which is not practical and has been included here solely for the purpose of setting an upper limit on achievable performance.

\subsection{Performance of the Cold-diffusion Model at Different Continuation Heights} 
\indent In this experiment, we generated diversified training data composed of four equal segments, each containing 5,500 samples with a density contrast of $-1$ g/cc, $-0.5$ g/cc, $0.5$ g/cc and $1$ g/cc, respectively. The cold-diffusion model is trained with 40 height steps, with a step size of $80\,$m, and evaluated for the downward continuation from the different continuation heights. The aim of this experiment is to assess the precision of the cold-diffusion model in continuing data from different heights using only one trained network. From Figure~\ref{fig10}, we observe that the cold-diffusion network can reconstruct data from different continuation heights within the RMSE of 0.15, along with direct reconstruction, demonstrating a superior performance over the stabilized DC for higher continuation heights. This study demonstrates that despite training multiple U-Nets to visualize the observed data at different depths, the proposed method can achieve reliable data reconstruction using only one trained network, thus making this approach computationally more efficient.

\subsection{Ablation Studies}
\begin{figure}[!t]
	\centering
	\includegraphics[width=4.25in]{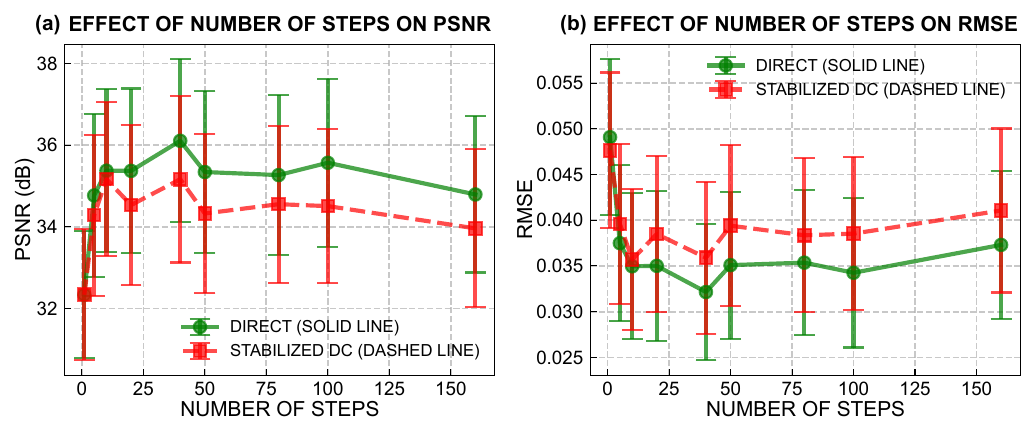}
	\caption{Effect of the number of diffusion steps on the performance of the cold-diffusion network for downward continuation.}
	\label{fig11}
\end{figure}

\indent In this experiment, the field data are gridded with a grid spacing of $400\,$m, and the continuation height is $800\,$m to simulate the smoothing similar to the synthetic data at $50\,$m spacing and $100\,$m continuation height. Figure~\ref{fig11} 
illustrates the performance of cold-diffusion on the field gravity data in terms of PSNR and RMSE across different number of time steps. We observed that the cold-diffusion model achieved convergent performance at approximately 20 time-steps, with minimal variation as the number of time-steps increased.

\subsection{Limitations of the Cold-Diffusion Approach}
\indent The downward data has been scaled to the range $[-1, 1]$, ensuring that the network is trained to produce outputs within this range. Rescaling the data to fit within the range is feasible if the minimum and maximum values of the reconstructed data are known. These values can be approximated based on the findings from a prior drill-hole study conducted in the survey area.

\section{Conclusion}
\label{Conclusion}
\indent We showed that the cold-diffusion network has a significant potential for stable downward continuation of gravity data in practical applications. 
Notably, the upward continuation process employs an exponential kernel in the Fourier domain, enabling us to formulate downward continuation as a concurrent deconvolution problem. Leveraging this insight, we adapted a cold-diffusion-based model to address the downward continuation by solving concurrent deconvolution problems.
The conventional techniques to address the downward continuation problem are prominently the regularization-based method,  whose effectiveness is sensitive to the choice of the regularization parameter. 
Deep-learning-based techniques such as U-Net eliminate the need for parameter tuning during inference. However, our experiments reveal that the U-Net-based technique is highly susceptible to even minimal levels of correlated noise, which is realistic in the case of geophysical data. 
Our technique eliminates the need for parameter tuning during inference and yields quantitatively equivalent performance as Oracle Tikhonov in the presence of correlated noise. Additionally, the proposed framework facilitates data visualization at multiple depths using a single trained network, significantly improving computational efficiency over training multiple U-Net for each continuation depth.
Our method can also be extended to the downward continuation of magnetic data, which has potential application in unexploded ordnance (UXO) detection \cite{Li2012}. This research also contributes to a better interpretation of gravity data collected through UAVs, which often appear smoothed out due to the increased distance between the observation plane and the source compared to ground-based surveys.

\begin{appendix}
Let $G$ and $R$ represent the ground truth and reconstructed data, respectively. Define RANGE as the difference between maximum and minimum values of $G$. Let MSE denote the averaged squared distance between $G$ and $R$. Let $G_{ij}$ and $R_{ij}$, where $1 \leq i \leq M$ and $1 \leq j \leq N$, indicate the gravity anomaly values at each pixel for the ground truth and reconstructed images, respectively. The peak-signal-to-noise ratio (PSNR), relative error (RE), and root-mean-square-error (RMSE) are defined as:
\begin{eqnarray}
    \label{eqn-15}
    \text{PSNR}(G, R) &=& 10 ~\text{log}_{10}\frac{(\text{RANGE})^2}{\text{MSE}}.\\
    \label{eqn-16}
	\text{RE}(G, R) &=& \frac{||G - R||_2}{||G||_2}.\\
	\label{eqn-17}
	\text{RMSE}(G, R) &=& \sqrt{\frac{1}{M\times N} \sum_{i = 1}^{M} \sum_{j = 1}^{N}(G_{ij} - R_{ij})^2}.
\end{eqnarray}

Let $\mu_G$ and $\mu_R$ represent the means of the ground-truth and reconstructed data, respectively; and $\sigma_G$ and $\sigma_R$ denote their respective variances. Additionally, define $c_1 = (k_1 L)^2$ and $c_2 = (k_2 L)^2$, where $L$ is the dynamic range, with $k_1 = 0.01$ and $k_2 = 0.03$ as the default values. Then, the structural similarity index (SSIM) is computed as
\begin{equation}
\label{eqn-18}
\text{SSIM}(G, R) = \frac{(2 \mu_G \mu_R + c_1)(2 \sigma_{GR} + c_2)}{(\mu_G^2+ \mu_R^2+c_1)(\sigma_G^2 + \sigma_R^2 + c_2)}.
\end{equation}
\end{appendix}

\section*{Acknowledgments}
\indent The authors would like to thank the Geological Survey of India (GSI) for providing the field gravity data. The first author would also like to  acknowledge the use of Microsoft Copilot for refining the language and presentation in this manuscript. Adarsh Jain gratefully acknowledges the support of the Axis Bank Centre for Mathematics and Computing (ABCMC) and the Indian Institute of Science (IISc) through a PhD fellowship.

\end{document}

\title{Cold-Diffusion Driven Downward Continuation of Gravity Data\\{\textit{Supplementary Material}}}

\author{Adarsh Jain, Pawan Bharadwaj and  Chandra Sekhar Seelamantula, \IEEEmembership{Senior Member, IEEE},
\thanks{
Adarsh Jain is with the IISc Mathematics Initiative (IMI), Department of Mathematics, Indian Institute of Science, Bengaluru, Karnataka 560012, India (e-mail: adarshjain1@iisc.ac.in). 

Pawan Bharadwaj is with the Centre for Earth Sciences, Indian Institute of Science, Bengaluru, Karnataka 560012, India (e-mail: pawan@iisc.ac.in).

Chandra Sekhar Seelamantula is with the Department of Electrical Engineering, Indian Institute of Science, Bengaluru, Karnataka 560012, India (e-mail: css@iisc.ac.in).}}

\markboth{Preprint}%
{}

\maketitle

\section{Introduction}
\begin{figure*}[t]
	\centering
	\includegraphics[width=6in]{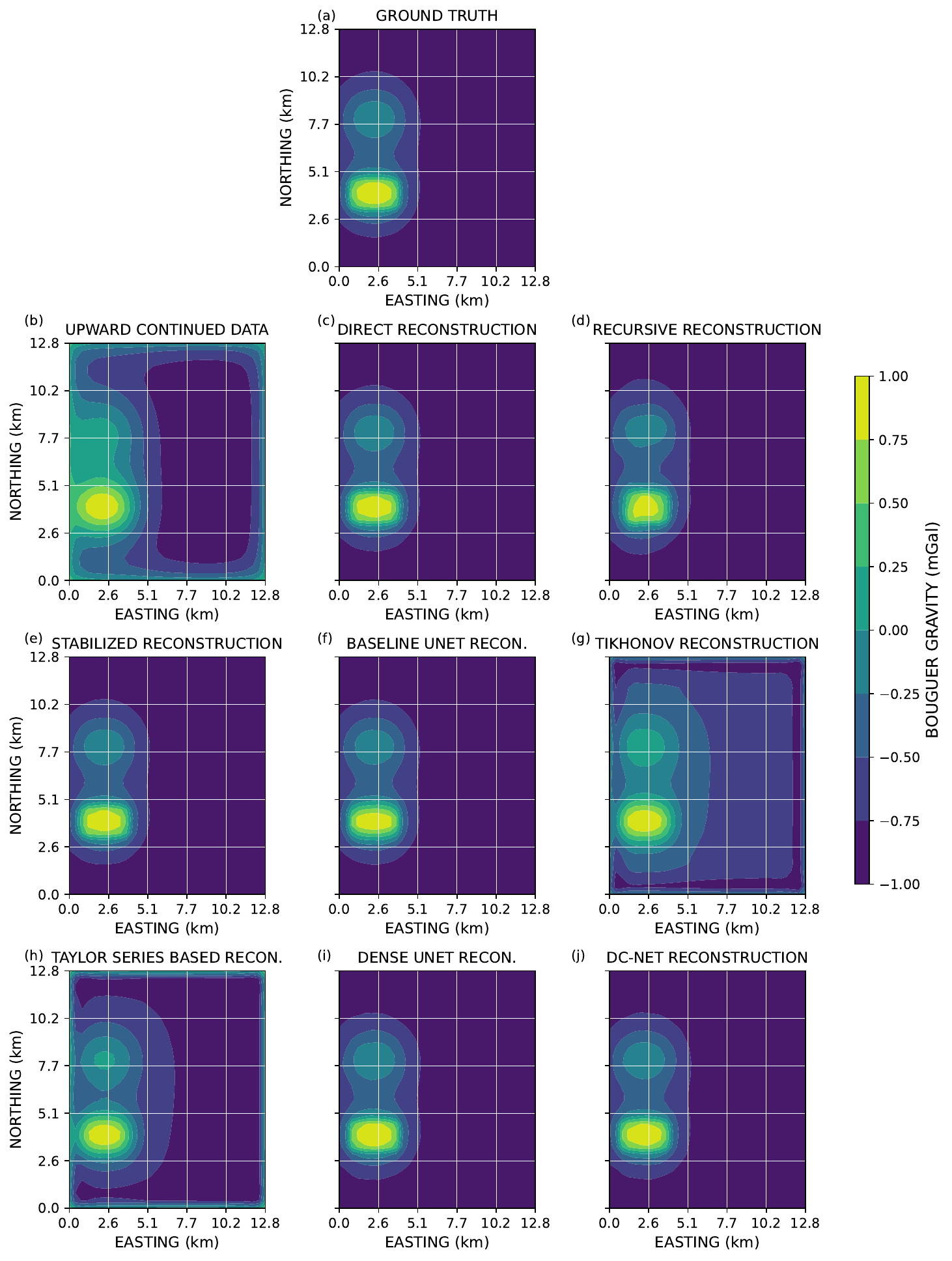}
	\caption{Qualitative assessment of various DC techniques on the synthetic data. The images have been normalized to the range of [$-1, 1$] for the purpose of visualization.}
	\label{fig7}
\end{figure*}

\begin{figure*}[t]
	\centering
	\includegraphics[width=6in]{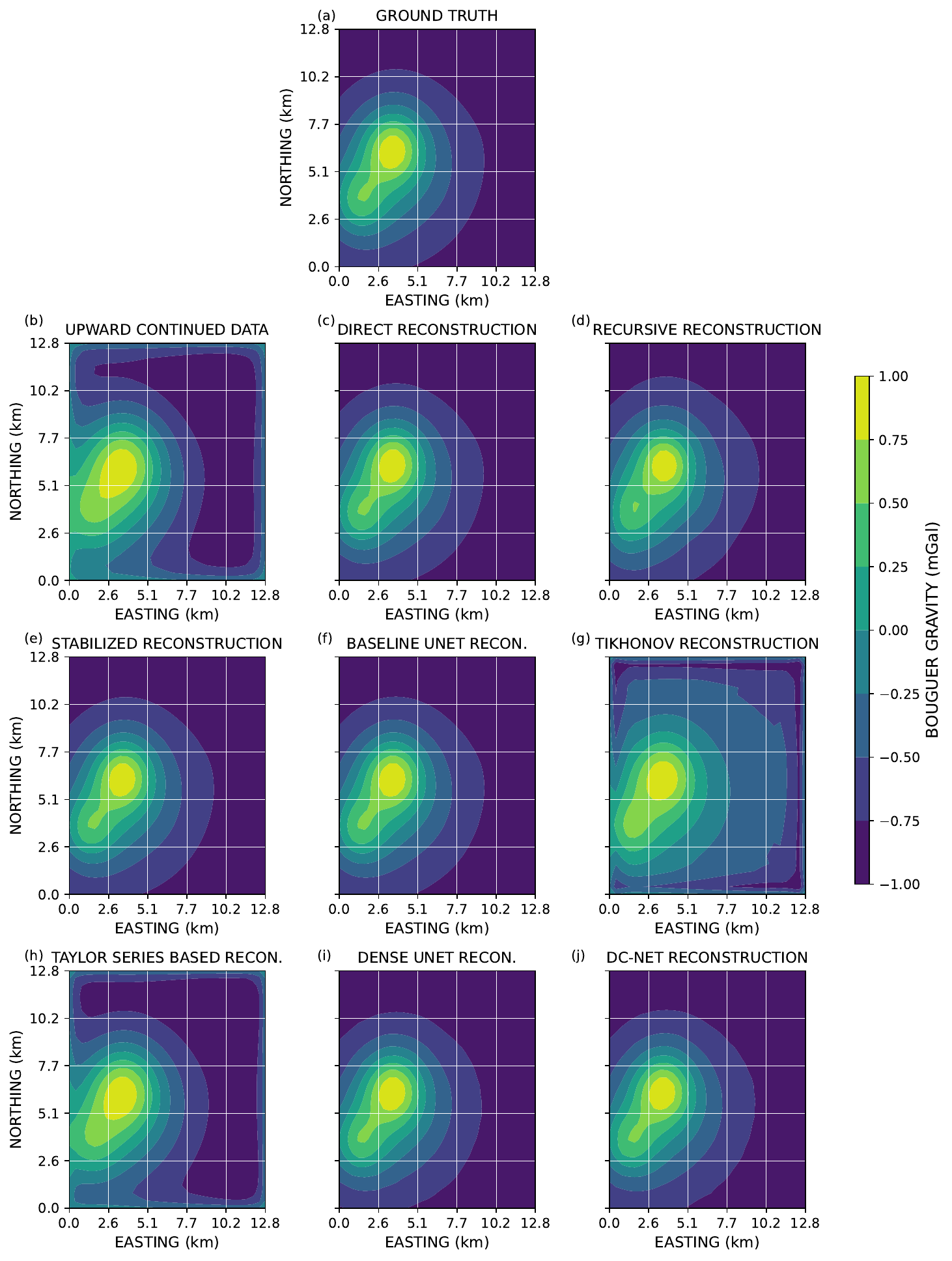}
	\caption{Qualitative assessment of various DC techniques on the synthetic data.}
	\label{fig8}
\end{figure*}

\begin{figure*}[t]
	\centering
	\includegraphics[width=6in]{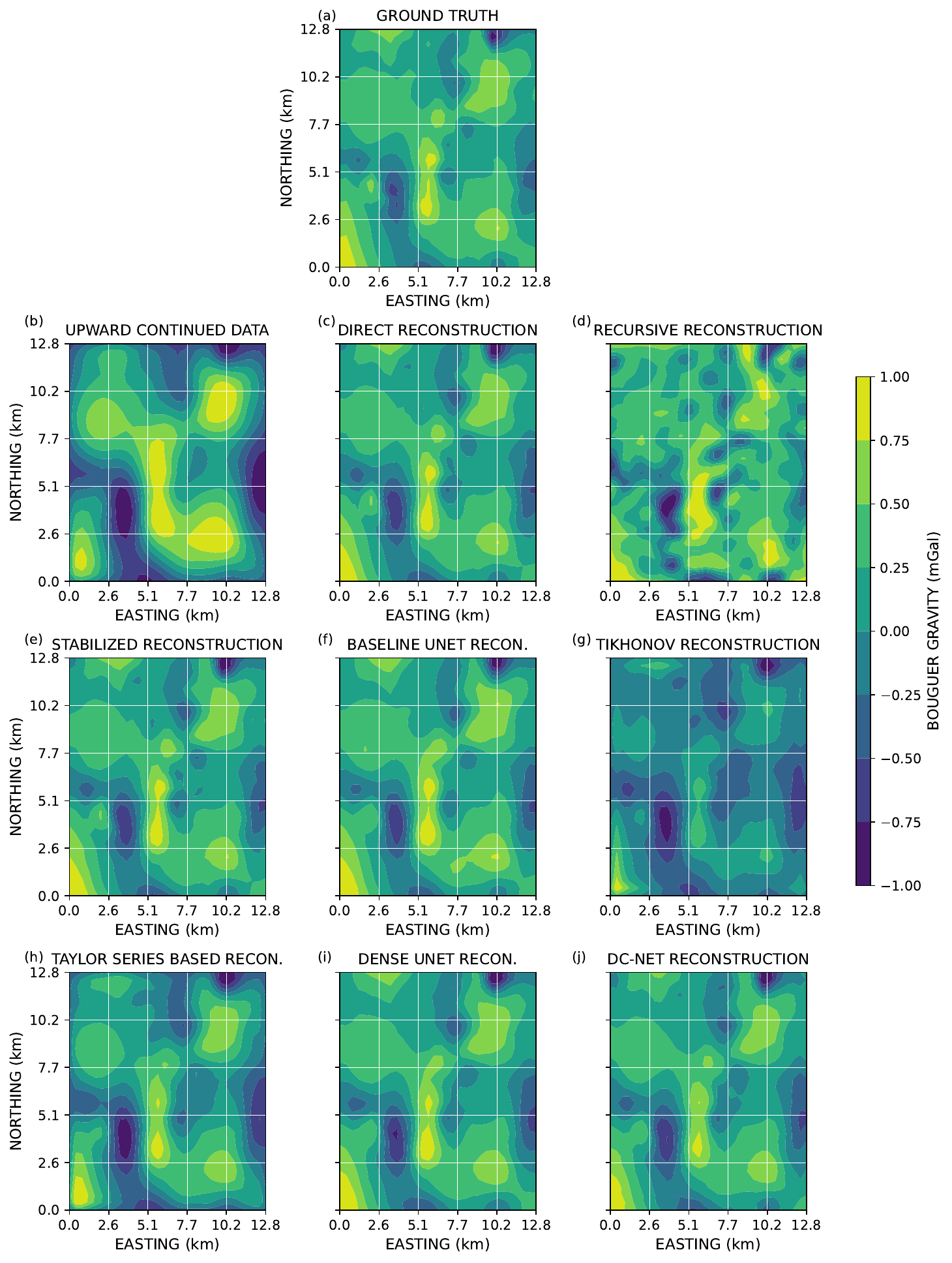}
	\caption{Qualitative assessment of various DC techniques on the field data.}
	\label{fig9}
\end{figure*}

\begin{figure*}[t]
	\centering
	\includegraphics[width=6in]{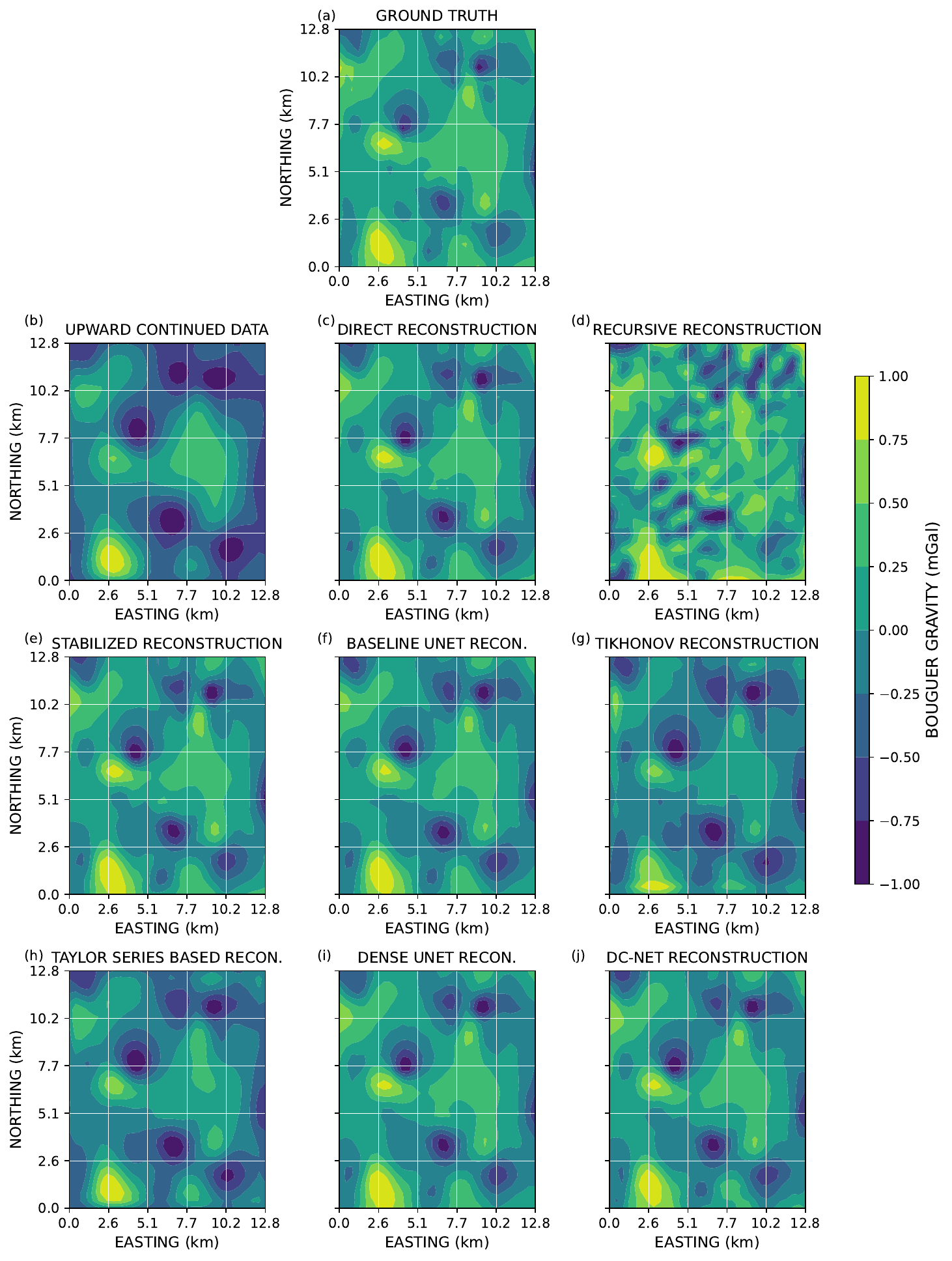}
	\caption{Qualitative assessment of various DC techniques on the field data.}
	\label{fig10}
\end{figure*}

\begin{figure*}[t]
	\centering
	\includegraphics[width=6in]{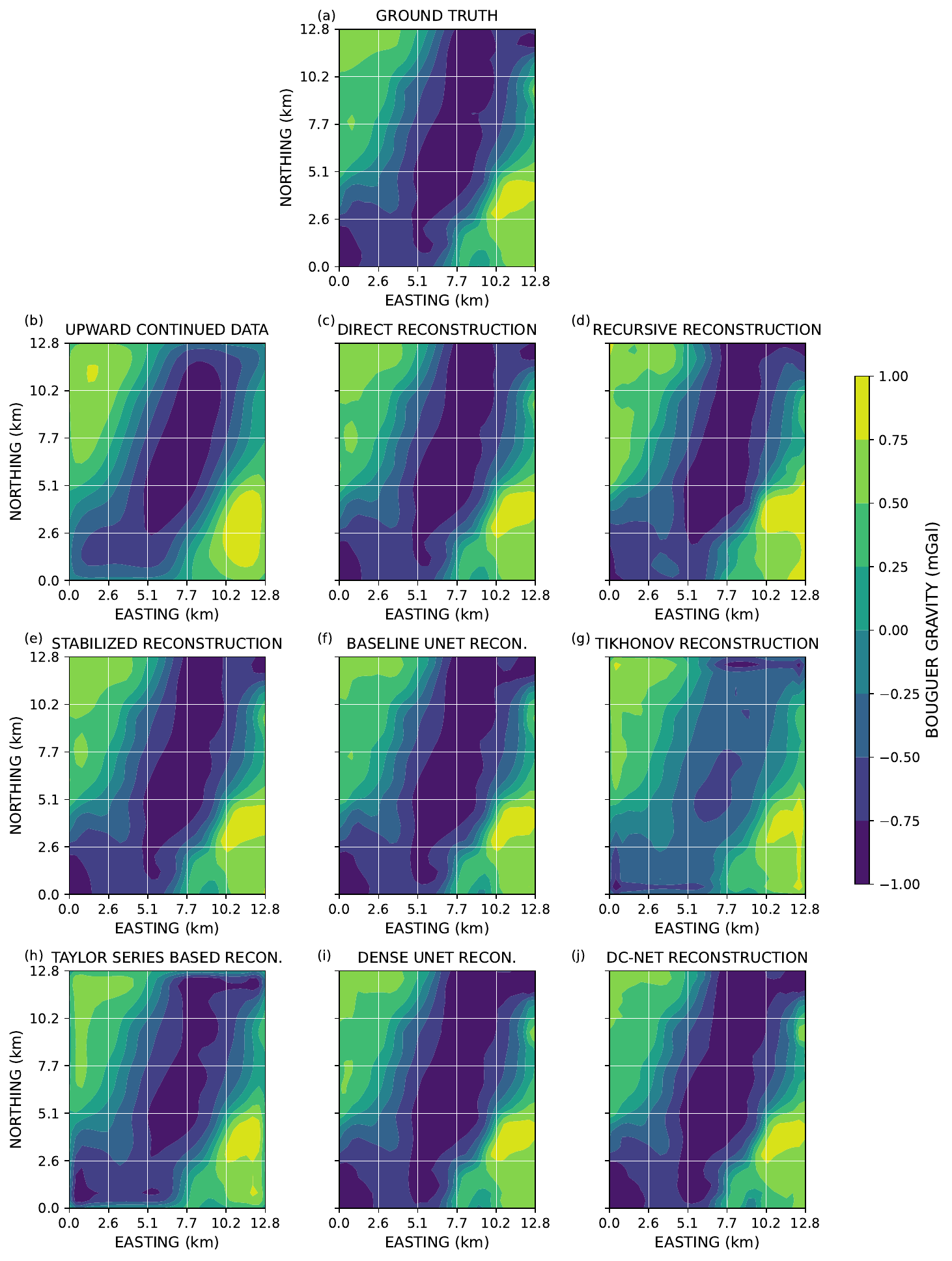}
	\caption{Qualitative assessment of various DC techniques on the field data.}
	\label{fig11}
\end{figure*}

\begin{figure*}[t]
	\centering
	\includegraphics[width=6in]{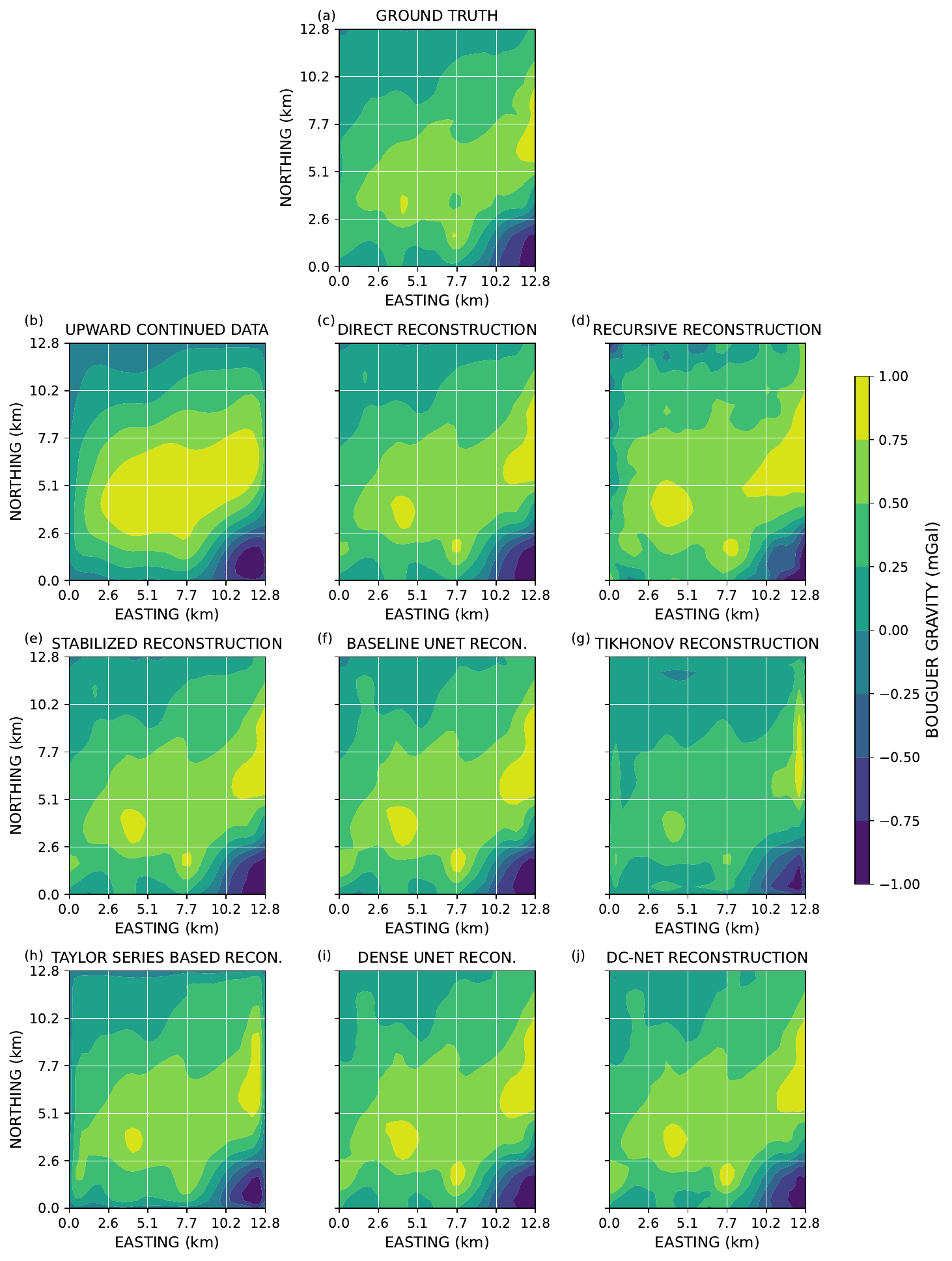}
	\caption{Qualitative assessment of various DC techniques on the field data.}
	\label{fig12}
\end{figure*}

\begin{figure*}[t]
	\centering
	\includegraphics[width=6in]{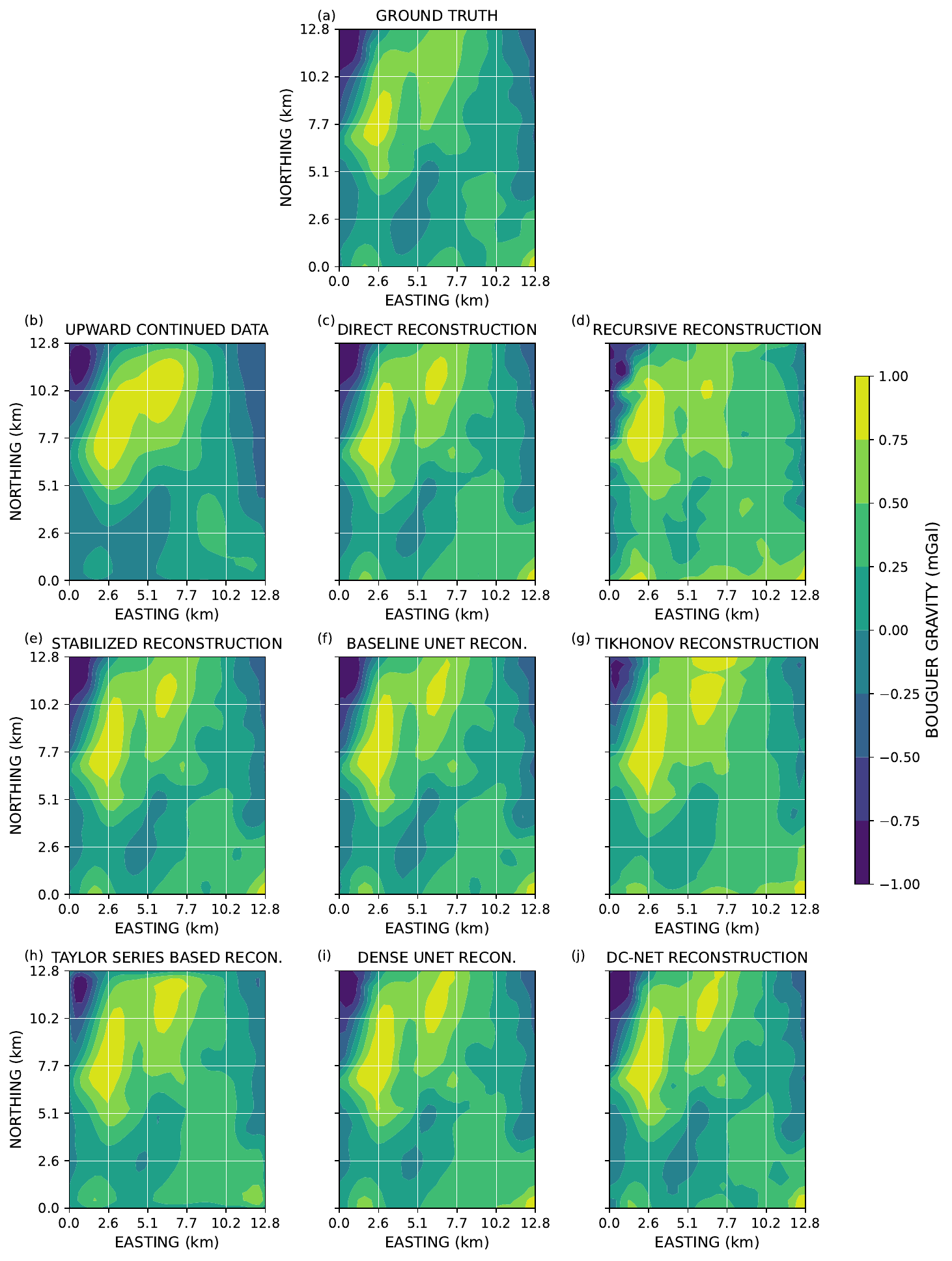}
	\caption{Qualitative assessment of various DC techniques on the field data.}
	\label{fig13}
\end{figure*}

\begin{figure*}[t]
	\centering
	\includegraphics[width=6in]{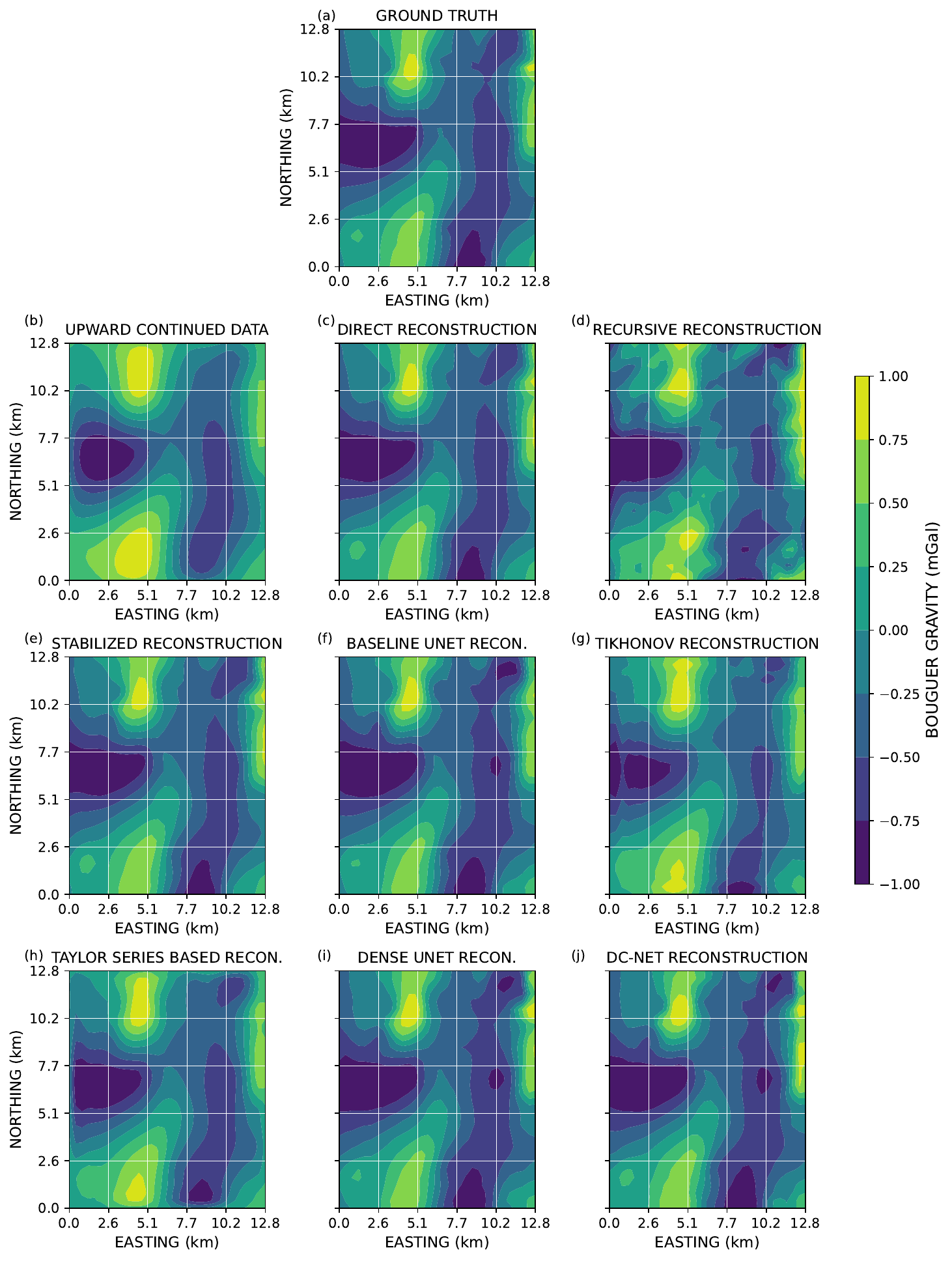}
	\caption{Qualitative assessment of various DC techniques on the field data.}
	\label{fig14}
\end{figure*}

In this document, we present additional qualitative analysis comparing our approach with current state-of-the-art methods. Additionally, we showcase the results for field gravity data using open-access datasets. Figures~\ref{fig7} and \ref{fig8} of this document illustrate the qualitative comparison of the proposed approach using synthetic data. Figures~\ref{fig9} and \ref{fig10} of this document present the results on the field data, with detailed information about this dataset provided in the paper.\\
\indent The open-access field gravity data utilized for the additional experiments is sourced from the \url{https://aikosh.indiaai.gov.in} website. This gravity data covers regions from Andhra Pradesh and Karnataka, India. The residual data was computed using the first-order surface fitting residualization method. For the current experimentation, we selected four patches of data ranging from 625000 to 640500 Easting and 1525000 to 1588500 Northing. The grid spacing for this gravity data is 500 meters. Figures \ref{fig11}, \ref{fig12}, \ref{fig13}, and \ref{fig14} of this document illustrate the performance of the proposed technique compared to existing state-of-the-art techniques for downward continuation corresponding to four different ground-truth images.


